\documentclass[12pt]{article}
\usepackage[english]{babel}
\usepackage[utf8]{inputenc}
\usepackage[T1]{fontenc}
\usepackage{authblk}
\usepackage{graphicx}
\usepackage{amsmath,mathrsfs,dsfont}
\usepackage{amsthm}
\usepackage[all]{xy}
\usepackage{graphics}
\usepackage{tensor}
\usepackage{verbatim}
\usepackage{extarrows}
\usepackage{nicefrac}
\usepackage{amsfonts}
\usepackage{amssymb,latexsym}
\usepackage{color}
\usepackage{mathrsfs,dsfont}
\usepackage{appendix}
\usepackage{slashed}
\usepackage{cite}
\usepackage{comment}

\newcommand{\be}{\begin{equation}}
\newcommand{\ee}{\end{equation}}
\newcommand{\bea}{\begin{eqnarray}}
\newcommand{\eea}{\end{eqnarray}}
\newcommand{\nn}{\nonumber}

\definecolor{verde}{cmyk}{.83,.21,1,.08}

\begin{document}

\title{Non-commutative $AdS_2/CFT_1$ duality:
 the case of massive and interacting scalar fields}
\author[1]{Felipe Rodrigues de Almeida\thanks{veliusrag@gmail.com}}
\author[1,2]{Aleksandr Pinzul\thanks{aleksandr.pinzul@gmail.com}}
\author[3]{Allen Stern\thanks{astern@ua.edu}}
\affil[1]{Universidade de Bras\'{\i}lia Instituto de F\'{\i}sica 70910-900, Bras\'{\i}lia, DF, Brasil}
\affil[2]{International Center of Physics C.P. 04667, Brasilia, DF, Brazil}
\affil[3]{Department of Physics, University of Alabama, Tuscaloosa, Alabama 35487, USA}
\date{}
\maketitle

\begin{abstract}
We continue the study of the nocommutative $AdS_2 / CFT_1$ correspondence. We extend our previous results obtained for a  free massless scalar field to the case of a massive scalar field.  Both the free and interacting cases are considered. For both cases it is confirmed that to the leading order in noncommutative corrections the 2- and 3-point correlation functions have the form that is assumed by some (yet unspecified) dual $CFT$. We also argue that there does not exist a map which connects the commutative model to its non-commutative counterpart,  and therefore  the conformal behaviour of the noncommutative correlators is a non-trivial result.
\end{abstract}

\section{Introduction}
In a recent paper \cite{Pinzul:2017wch} aspects of the $AdS_2 / CFT_1$ correspondence were studied in a non-commutative setting, namely when the geometry on the gravity side of the correspondence is replaced by the non-commutative version of two-dimensional anti-de Sitter space ($AdS_2$). The motivation for making the $AdS_2$ space  non-commutative is to include some quantum gravitational corrections, since there is a general belief (supported by multiple arguments) \cite{Doplicher:1994tu} that the quasiclassical regime of quantum gravity should appear as a quantum field theory on some non-commutative background. The  introduction of noncommutativity  on the   $AdS_2$  background can be made unique by demanding that it preserves the $SO(2,1)$ isometry group.\cite{Ho:2000fy},\cite{Ho:2000br},\cite{Fakhri:2011zz},\cite{Jurman:2013ota},\cite{Stern:2014aqa},\cite{Chaney:2015ktw} In \cite{Pinzul:2017wch}, analogues of  $AdS_2$  Killing vectors generating $SO(2,1)$ were constructed on the noncommutative space, denoted by $ncAdS_2$.  Such  $ncAdS_2$ Killing vectors are deformations of the $AdS_2$  Killing vectors $K_\mu$, yet, due to the unique introduction of noncommutativity, they generate the {\it undeformed } isometry group, i.e., $SO(2,1)$. These deformations were shown to vanish in the boundary limit, and so the  $ncAdS_2$ Killing vectors asymptotically tend towards $K_\mu$. In other words, $ncAdS_2$, is asymptotically $AdS_2$. The $AdS/CFT$ correspondence principle posits a weak/strong duality between the quantum gravity in the bulk of an asymptotically $AdS$ space and a conformal field theory (CFT) on the  boundary of this space.  Baring the known difficulties of the correspondence principle for two dimensional anti-de Sitter space (see for example, \cite{Strominger:1998yg},\cite{Maldacena:2016hyu}), the result that $ncAdS_2$, is asymptotically $AdS_2$ opens up the possibility of a dual conformal field theory on the boundary.

In the usual prescription for the $AdS/CFT$ correspondence principle, the connected correlation functions for operators ${\cal O}$ spanning the $CFT$ are generated by the on-shell field theory action on the corresponding asymptotically $AdS$ space, and the boundary values $\phi_0$ of the fields are sources associated with ${\cal O}$. The article \cite{Pinzul:2017wch}  specialized to the case of a single ``free'' massless scalar field on $ncAdS^2$. This provided a particularly simple example, in part because of the fact that solutions to the field equation on $AdS_2$ are regular at the boundary, i.e., $|\phi_0| <\infty$. Perturbative theory was utilized to compute the leading corrections to two-point correlation function of operators on the boundary induced by the bulk-to-boundary and bulk-to-bulk propagators. It was found that these corrections have the same form as the commutative result, and thus preserve the $SO(2,1)$ conformal symmetry at the boundary. This result is consistent with the isometry preserving construction of noncommutativity.\footnote{This is in contrast to the introduction of $\kappa$-spacetime in the bulk which resulted in a deformed conformal symmetry on the boundary\cite{Gupta:2017xex}.}

In this article we extend the work of \cite{Pinzul:2017wch} to include the case of both a free and interacting massive scalar field on $ncAdS_2$. We obtain an exact result for the leading noncommutative correction to the two-point correlation function on boundary. As with the massless limit, it preserves the $SO(2,1)$  conformal symmetry. In the interacting case, we add a cubic term to the action, and from it we obtain an integral expression for the leading order noncommutative correction to the three point correlation function on the conformal boundary.  It too is seen to be consistent with $SO(2,1)$ conformal symmetry. In this article we also argue that there does not exist a map connecting the commutative Killing vectors to their noncommutative analogues, an thus that the $ncAdS^2$ theory cannot be trivially related to its commutative counterpart.

The outline of this article is the following: After briefly reviewing the correspondence principle for free massive scalar field theory on $AdS^2$ in section two, we quantize the background space to get $ncAdS^2$, and derive the leading order noncommutative correction to the two point correlation function on the conformal boundary. We introduce a cubic interaction in the scalar field  action on $ncAdS^2$, and compute corrections to the three point correlation function in section three. In the  conclusion (section four), we briefly summarize the obtained results and discuss some possible directions for future developments. Some technical results are collected in the four appendices. In appendix A, we give some useful expressions for the asymptotic form of the relevant Green functions. In appendix B, we give the map between the noncommutative versions of canonical coordinates and Fefferman-Graham coordinates. In appendix C we argue that there does not exist a trivial map connecting the noncommutative and commutative field theory. In appendix D we show that the on-shell boundary action does not pick up noncommutative corrections.

\section{Free massive scalar field}
\setcounter{equation}{0}
\subsection{Commutative case}

We start by briefly reviewing the definition of the 2-dimensional anti-deSitter space, $AdS_2$. One should distinguish between Lorentzian and Euclidean $AdS$. Lorentzian $AdS$, which is standardly denoted by $AdS_2$, can be defined using the global embedding coordinates, $X_\mu$, in $\mathbb{R}^{2,1}$ with the signature $(-,+,-)$ upon imposing the constraint:
\bea\label{AdS}
g_{\mu\nu}X^\mu X^\nu = - \ell_0^2 \ ,\text{where}\ \ell_0^2 \in \mathbb{R}_+ \ \text{and}\ g_{\mu\nu}=\mathrm{diag}(-1,1,-1)\ .
\eea
To avoid the closed time-like curves, one passes to a covering space.  (See for example \cite{Natsuume:2014sfa}.)

The Euclidean anti-deSitter space, which is standardly denoted by $EAdS_2$, is defined by the analogous embedding in three-dimensional Minkowski space $\mathbb{R}^{2,1}$, but now with the signature  $(+,+,-)$,\footnote{This is equivalent to the Wick rotation in global coordinates.}
\bea\label{EAdS}
g^{E}_{\mu\nu}X^\mu X^\nu = - \ell_0^2 \ ,\ \ell_0^2 \in \mathbb{R}_+ \ ,
\eea
with  indices raised and lowered using the  metric tensor  $g^{E}={\rm diag}(1,1,-1)$. In both cases $\ell_0$ is the scale parameter. Equivalently, (\ref{AdS}) [(\ref{EAdS})] represents $EAdS_2$ [$AdS_2$] for $\ell_0$ imaginary. In this article we restrict to the case of the Euclidean version and will often refer to it as just $AdS_2$. As it is clear from (\ref{EAdS}), in this case the constraint describes a double-sheeted hyperboloid.

There are two convenient choices for the coordinate charts of the  lower hyperboloid, $X^2 <0$, (and the same for the upper one with some sign changes).  They are the canonical coordinates, $(x,y)$, and the Fefferman-Graham (FG) coordinates, $(t,z)$. The canonical coordinatization was defined in \cite{Pinzul:2017wch} by
\bea\label{corrd_canon}
X^0&=&-y\ , \cr&&\cr
X^1&=&-\frac{1}{2\ell_0} y^2 e^{-x} + \ell_0 \sinh x\ , \cr&&\cr
X^2&=&-\frac{1}{2\ell_0} y^2 e^{-x} - \ell_0 \cosh x\ ,
\eea
where $-\infty < x,y < \infty$.

The Fefferman-Graham coordinatization \cite{Fefferman} of the lower hyperboloid is given by
\bea\label{corrd_FG}
X^0&=&-\frac{\ell_0 t}z\ , \cr&&\cr
X^1&=&-\frac{\ell_0 }2\,\Bigl(z+\frac{t^2-1}z\Bigr)\ , \cr&&\cr
X^2&=&-\frac{\ell_0 }2\,\Bigl(z+\frac{t^2+1}z\Bigr)\ ,
\eea
where the coordinates span the half-plane, $z> 0$, $-\infty<t<\infty$ and $z\rightarrow +0$ corresponds to approaching the boundary.

The relation between two coordinate systems is given by
\bea\label{xyFG}
\left\{
\begin{array}{l}
  x=-\ln z \\
  y=\ell_0 \frac{t}{z}
\end{array}
\right.
\Leftrightarrow
\left\{
\begin{array}{l}
  t= \frac{1}{\ell_0}y e^{-x}\\
  z= e^{-x}
\end{array}
\right. \ .
\eea
We will see that the canonical coordinates are essential to quantize the geometry.  On the other hand, for the case of the `classical' geometry we can work directly in terms of  the Fefferman-Graham coordinates. The metric tensor when expressed in FG coordinates is given by
\be
ds^2=\frac{\ell_0^2}{z^2}\;\Bigl( dz^2+ dt^2\Bigr) \ ,
\ee
while the Euclidean Laplace operator is
\be\label{cmtvlplcin}
\Delta^{(0)} = \frac{z^2}{\ell_0^2}(\partial^2_z + \partial_t^2)=:\frac{1}{\ell_0^2}{\cal L}^{(0)}\ .
\ee
${\cal L}^{(0)}$ can be expressed in terms of the three Killing vectors $K^\mu$, $\mu=0,1,2$,  for  $AdS_2$, satisfying $so(2,1)$ Lie algebra commutation relations  $[K^\mu, K^\nu]=\epsilon^{\mu\nu\rho}K_{\rho}$. In terms of the FG coordinates they are given by
\be\label{KlngfG}
K^-=-\partial_t\quad ,\quad K^0=-t\partial_t-z\partial_z\quad ,\quad K^+=(z^2- t ^2)
\,\partial_t -2 z t \,\partial_z \;,
\ee
where $K^\pm=K^2\pm K^1$.  (For the expressions in terms of the canonical coordinates, see (\ref{Killings}).) It can be checked that (\ref{cmtvlplcin}) is the quadratic Casimir of the $so(2,1)$ Lie algebra
\be\label{Klngcmtvsffe}
{\cal L}^{(0)}= K^\mu K_\mu \ .
\ee
As we will see, this has a straightforward generalization in the non-commutative theory.

When expressed in terms of Fefferman-Graham coordinates the action for a free real massive scalar field  $\Phi^{(0)}$  in a Euclidean $AdS_2$ background is
\bea\label{clmsfa}
S[\Phi^{(0)}]&=&\frac 12\int_{ {{R}}\times  {{R}}_+} dt dz\,\,\Bigl\{( \partial_z\Phi^{(0)} )^2 \,+\,(\partial_t\Phi^{(0)})^2  \,+\,\Bigl(\frac{m_0\ell_0}z\Bigr)^2{\Phi^{(0)}}^2 \Bigr\}\ ,\label{cmtvactn}
\eea
where $m_0$ is the mass. The $(0)$ superscript on the field denotes the commutative theory. The scalar field equation  resulting from (\ref{cmtvactn}) is
\be\label{cmtvsffe}
{\cal L}^{(0)}\Phi^{(0)}=({m_0\ell_0})^2{\Phi^{(0)}} \ .
\ee
Near the boundary, which is located at $z= +0$,\footnote{Below, ``$z=0$'' should be understood as ``$z= +0$''.} the dominant solution behave as  (for a general discussion of asymptotic behavior, see for example \cite{Ammon:2015wua}.)
\be\label{smtptccmtvnsr}
\Phi^{(0)}(z,t)\rightarrow z^{\Delta_-}\phi_0(t)\ ,
\ee
where $\Delta_\pm =\frac{1}{2}\pm\nu $, $\nu=\sqrt{\frac{1}{4} +(m_0\ell_0)^2}$. $m_0$ is not necessarily real, although it satisfies the Breitenlohner-Freedman bound, $(m_0\ell_0)^2> -\frac{1}{4}$.\cite{Breitenlohner:1982bm,Breitenlohner:1982jf} This bound comes about from the requirement of the absence of the normalizable negative energy states, i.e. instabilities of the theory. We shall assume that the boundary function $\phi_0(t)=\lim\limits_{z\rightarrow 0}\,z^{-\Delta_-}\,\Phi^{(0)}(z,t)$ is nonvanishing. Then from (\ref{smtptccmtvnsr}), $\Phi^{(0)}$ is singular in the boundary limit when $(m_0\ell_0)^2>0$.

Away from the boundary, regular   solutions to (\ref{cmtvsffe}) can be expressed in terms of $ \phi_0(t)$ using the boundary-to-bulk propagator (\ref{Bulk2boundary})
\be\label{frstrdrntsln}
\Phi^{(0)}(z,t)= \int_{ {{R}}} dt'\, K(z,t;t')\,\phi_0 (t')\ , \qquad  K(z,t;t')=  C_{\Delta_+}\biggl(\frac{z}{z^2+(t-t')^2}\biggr)^{\Delta_+}\ ,
\ee
where $ C_{\Delta_+}=\frac {\Gamma(\Delta_+)}{\sqrt{\pi}\,\Gamma(\nu)}$. We denote such solutions by $\Phi^{(0)}_{sol}[\phi_0]$. They are regular  for all  $z\geq 0$.

Substituting the solutions (\ref{frstrdrntsln}) back into  (\ref{clmsfa}) gives  the  on-shell action, which is a  boundary term:
\bea
S\Bigl[\Phi^{(0)}_{sol}[\phi_0]\Bigr]&=&-\frac 12\int_{ {{R}}} dt \;\Phi^{(0)}_{sol}[\phi_0] \; \partial_z\Phi^{(0)}_{sol}[\phi_0] \Big|_{z=0}\label{cmtvBndTurm}\\
&&\cr &=&-\frac 12\int_{ {{R}}} dt  \int_{ {{R}}} dt' \int_{ {{R}}} dt''\;K(z,t;t')\, \partial_z K(z,t;t'') \Big|_{z=0}\;\phi_0(t') \phi_0 (t'')\label{six} \\
&&\cr&=&-\frac {\Delta_+\Gamma(\Delta_+)}{2\sqrt{\pi}\,\Gamma(\nu)}\int_{ {{R}}} dt' \int_{ {{R}}} dt''\,\frac{\phi_0(t')\phi_0(t'') }{|t'-t''|^{2\Delta_+}}\ ,\label{zrthrdrosa}
\eea
where we used the boundary value (\ref{AsymptoticsKK}) and the result is valid for any $t'\ne t''$ on the boundary.

The standard prescription for the  $AdS/CFT$ correspondence is to identify the on-shell action with the generating functional of the $n-$point connected correlation functions for operators ${\cal O}$ defined on the boundary.\cite{Witten:1998qj} Here $\phi_0$ is treated as the source for ${\cal O}$. For a 2-dimensional theory in the bulk, both ${\cal O}$ and  $\phi_0$ are  functions of only one coordinate, the time $t$. The $n-$point connected correlation functions are thus
\be\label{adscft}
<{\cal O}(t_1)\cdots{\cal O}(t_n)>^{(0)}=\frac{\delta^n S[\Phi^{(0)}_{sol}[\phi_0]]}{\delta \phi_0(t_1)\cdots\delta \phi_0(t_n)}\bigg|_{\phi_0=0} \ ,
\ee
the ${(0)}$ superscript again  indicating that this is the commutative theory. The two-point function for the example of the massive scalar is then
\be\label{2ptfncmtv}
<{\cal O}(t){\cal O}(t ')>^{(0)}\;=\;\,\frac{\gamma^{(0)}(\Delta_+) }{|t-t'|^{2\Delta_+}}\ ,\qquad \gamma^{(0)}(\Delta_+)=\,-\frac {\Delta_+\Gamma(\Delta_+)}{\sqrt{\pi}\,\Gamma(\nu)} \ .
\ee
This is exactly the form expected for the 2-point function in conformal theory, the conformal dimension being $\Delta_+$.

\subsection{Non-commutative case}\label{SectionNCfree}

Now we want to adopt the previous construction to the suitable non-commutative deformation of the Euclidean $AdS_2$. In so doing, we  set the stage for the much less trivial case of an interacting field (see the section \ref{SectionNCint}), as well as generalize the treatment in \cite{Pinzul:2017wch} where we restricted to the case of a massless field. If we are to have any hope of preserving the conformal symmetry at the boundary, we should quantize the $AdS$ geometry in the way that preserves the isometries of the commutative counterpart. An analogous example of such a deformation is the fuzzy sphere, where the $SO(3)$ rotation symmetry is preserved upon quantization.\cite{Madore:1991bw},\cite{Grosse:1994ed},\cite{CarowWatamura:1998jn}, \cite{Alexanian:2000uz},\cite{Dolan:2001gn},\cite{Balachandran:2002jf} (see \cite{Alexanian:2000uz} for the discussion of the coherent states and the recovery of the commutative limit). The preservation of the symmetries (as opposed to their breaking or deformation) allows for many results to be obtained on a purely algebraic level. So, in our case we are aiming at the non-commutative version of $AdS_2$ ($ncAdS_2$), which preserves the full $so(2,1)$ isometry of $AdS_2$.  It is defined similarly to the case  of fuzzy sphere by three hermitian operators,  $\hat X^\mu$, here satisfying,\cite{Ho:2000fy},\cite{Ho:2000br},\cite{Fakhri:2011zz},\cite{Jurman:2013ota},\cite{Stern:2014aqa},\cite{Chaney:2015ktw}
\be\label{Ucldadscasimir}
\hat X^\mu \hat X_\mu=-\ell^2\mathds{1}\ ,
\ee  which is analogous to the $AdS$  constraint (\ref{EAdS}).
As before, raising/lowering of the indices is performed using the metric $g^E$ (\ref{EAdS}).  In addition to (\ref{Ucldadscasimir}),  $\hat X^\mu$ is a basis for the $so(2,1)$ algebra
\be\label{adstoocrs}
[\hat X^\mu,\hat X^\nu]=i\alpha\,\epsilon^{\mu\nu\rho}\hat X_\rho\ ,
\ee
where $\alpha $ and $\ell$ are two real parameters with units of length. The former is the noncomutativity parameter, while the latter can be interpreted as the length scale of $ncAdS_2$. $\mathds{1}$ denotes the identity operator. The commutative limit corresponds to $(\alpha,\ell)\rightarrow(0,\ell_0)$. It is easy to see that (\ref{Ucldadscasimir}) is the quadratic Casimir of the algebra generated by $\hat X^\mu$ subject to the relations (\ref{adstoocrs}). But now, in contrast with the finite dimensional representations of the algebra of a fuzzy sphere, due to the non-compactness of the $SO(2,1)$ group, we have to consider infinite unitary  representations. A detailed study of these representations, as well as the interplay between the Lorentzian and Euclidean cases, was performed in \cite{Pinzul:2017wch}.

The non-commutative generalization of the action (\ref{clmsfa}) for a free massive field is given by the standard form
\be\label{mslssclrfldactn}
S_{nc}[\hat \Phi]=-\frac{1}{2\ell}{\rm Tr}\,\Bigl\{[\hat X^\mu,\hat \Phi][\hat X_\mu,\hat \Phi]-(\alpha\ell m)^2 \hat\Phi^2 \Bigr\}\ ,
\ee
where $\hat\Phi$ is non-commutative field on $ncAdS$, Tr denotes a trace, and  $m$ is the mass of the scalar field.  We assume that $m\rightarrow m_0$ in the commutative limit in order to recover the previous  scalar field dynamics.

As was discussed in detail in \cite{Pinzul:2017wch}, the canonical coordinates (\ref{corrd_canon}) upon quantization satisfy the standard canonical relation
\be\label{canonic_comm}
[x,y]=i\alpha\mathds{1}\ .
\ee
(Here for convenience we use the same letters for the commutative and non-commutative coordinates.)
Then one can pass from the operator algebra generated by $\hat X^\mu$ to the algebra of their symbols ${\cal X}^\mu(x,y)$,  with the operator product replaced by the Moyal-Weyl star product, $\star$. (for a nice review, see \cite{Zachos:2000zh})  The Moyal-Weyl star product of any two symbols ${\cal F}$ and $ {\cal G}$ is defined in the standard way by
\be\label{dffstrprd}
[ {\cal F}\star  {\cal G} ](x,y) =  {\cal F}(x,y)\,\exp{\Bigl\{\,\frac{i\alpha}2 \,(\overleftarrow{ { \partial_ x }}\,
\overrightarrow{ {\partial_ y }}\,-\,\overleftarrow{ { \partial_ y }}\,
\overrightarrow{ {\partial_ x }}
)\,
\Bigr\}}\; {\cal G}(x,y)\ ,
\ee
while the trace on the algebra, Tr, becomes $ \frac 1{\alpha^2} \int_{ {{R}}^2} dxdy$. It was shown in  \cite{Pinzul:2017wch} that the symmetric ordering in the deformation of the canonical coordinatization (\ref{corrd_canon})
\bea\label{smnlsofXmu}
{\cal X}^0 &=&- y \ ,
\cr &&\cr  {\cal X}^1 &=&- \frac {1}{2\ell} \, y\star  e^{-x}\star y +\ell\sinh x\ ,
\cr &&\cr  {\cal X}^2 &=&- \frac {1}{2\ell} \, y\star e^{-x}\star y-\ell\cosh x\
\eea
leads to the correct star product  product realization of the defining relations, (\ref{Ucldadscasimir}) and (\ref{adstoocrs}), for $ncAdS^2$,
\bea
{\cal X}^\mu  \star {\cal X}_{ \mu}&=& - \ell^2 \ ,\\
&&\cr  [ {\cal X} ^\mu, {\cal X} ^\nu]_\star &=&i\alpha\,\epsilon^{\mu\nu\rho} {\cal X}_{\rho}\ ,
\eea
where $ [{\cal F}, {\cal G} ]_\star={\cal F}\star {\cal G} -{\cal G}\star {\cal F}$ is the star-commutator of any two functions ${\cal F}( x, y)$ and ${\cal G}( x, y)$ on the  Moyal-Weyl plane spanned by coordinates $(x,y)$.

Using this, the action (\ref{mslssclrfldactn}) can be trivially mapped to a functional on the Moyal-Weyl plane
\be\label{Sclrfldactn}
S_{nc}[\Phi]=-\frac 1{2\ell\alpha^2} \int_{ {{R}}^2} dxdy\,\Bigl\{[{\cal X}^\mu,\Phi]_\star\star [{\cal X}_{\mu},\Phi]_\star-(\alpha\ell m)^2 \Phi\star\Phi\Bigr\} \ ,
\ee
where $\Phi=\Phi (x,y)$ is  the  symbol of the field $\hat \Phi $. Varying the action with respect to $\Phi$, one gets the corresponding equation of motion
\be\label{fenblk}
{\cal L}\Phi:=- \frac 1{\alpha^2}[{\cal X}^\mu,[{\cal X}_{\mu},\Phi]_\star]_\star =(m\ell)^2\Phi  \ ,
\ee
where ${\cal L}$ is the noncommutative Laplace operator.

Alternatively, the action (\ref{Sclrfldactn}) and the equation of motion (\ref{fenblk}) can be written explicitly in terms of the canonical coordinates as \cite{Pinzul:2017wch}. For the former we get
\be\label{LntrmsofDltS}
S_{nc}[\Phi]=\frac 1{2\ell}\int_{ {{R}}^2} dxdy\,\biggl\{ \Bigl( y \Delta_y\Phi+ \partial_x S_y\Phi\Bigr)^2 \;+\;\Bigl(\frac {\alpha^2}4+{\ell^2}\Bigr)(\Delta_y\Phi)^2+ ( m \ell)^2\Phi^2 \biggr\} \ ,
\ee
up to boundary terms, where $\Delta_y$ and $S_y$ are the nonlocal operators
\bea\label{dfDltaS}
\Delta_y\Phi(x,y)&=&\frac{\Phi\Bigl(x,y+\frac {i\alpha}2\Bigr)-\Phi\Bigl(x,y-\frac {i\alpha}2\Bigr)}{i\alpha}\;=\; \frac 2\alpha\sin\Big(\frac{\alpha}2\partial_y\Bigr)\Phi(x,y)\ ,\cr &&\cr  S_y\Phi(x,y)&=&\frac{\Phi\Bigl(x,y+\frac {i\alpha}2\Bigr)+\Phi\Bigl(x,y-\frac {i\alpha}2\Bigr)}{2} \;=\; \cos\Big(\frac{\alpha}2\partial_y\Bigr)\Phi(x,y) \ ,
\eea
while for the latter
\be\label{nclplnxy}
{\cal L}\Phi=(  \Delta_yy+ \partial_x S_y )\,( y \Delta_y +\partial_x S_y)\,\Phi  \;+\;\Bigl(\frac {\alpha^2}4+{\ell^2}\Bigr)\, \Delta_y^2\Phi =(m\ell)^2\Phi\ .
\ee

In complete analogy to (\ref{Klngcmtvsffe}), ${\cal L}$ can also be written in terms of the noncommutative Killing vectors  $K^\mu_\star$. In terms of the canonical coordinates $(x,y)$, the non-commutative  Killing vectors are given by (\ref{ncKillings})
\bea\label{Mnxndy}
\quad K_\star^0&=&\partial_x\;,\qquad K_\star^-\;=\;-{\ell} \, e^{ x} \Delta_y\ ,\cr &&\cr
K_\star^+&=& \frac {e^{-x}} {\ell}\,\Biggl(2 y\,\partial_x S_y+\biggl(y^2+{\ell^2}+ \frac {\alpha^2}4(1-\partial_x^2)\biggr)\, \Delta_y\Biggr)\ .
\eea
They preserve the $so(2,1)$ Lie algebra commutation relations, $[K_\star^\mu, K_\star^\nu]=\epsilon^{\mu\nu\rho}K_{\star\rho}$. While $K_\star^0$ is identical to its commutative analogue $K^0$, $K_\star^\pm=K_\star^2\pm K_\star^1$ are deformations of $K^\pm$, which are given in  (\ref{Killings}). $K_\star^\pm\rightarrow K^\pm$ in the commutative limit. As in the commutative theory, the Laplace operator is the quadratic Casimir of the $so(2,1)$ Lie algebra
\be\label{Klngcmtvsffenc}
{\cal L}=K_\star^\mu { K_\star}_\mu\ .
\ee

We search for a perturbative solution to the field equations, with $\frac{\alpha^2}{\ell^2}$ being the perturbative parameter. For this we will use the method discussed in the Appendix \ref{Similar}. Namely, we will map the full non-commutative equation (\ref{fenblk}) into the commutative one (\ref{cmtvsffe}) plus some corrections. To achieve this, we construct an operator $U$, given explicitly up to order $\alpha^2$ in (\ref{Transffinal}), which maps the set of the commutative Killing vectors (\ref{Killings}) to the non-commutative ones (\ref{ncKillings}) modulo an additional term in $K_\star^+$. Taking into account that in both cases the Laplacian can be written in terms of their respective Killing vectors, (\ref{Klngcmtvsffe}) and (\ref{Klngcmtvsffenc}), the necessary map between the Laplacians is trivially found up to order $\alpha^2$ to be
\be\label{uclLuinv}
U{\cal L}U^{-1}=U K^\mu_\star K_{\star\mu}U^{-1} = {\cal L}^{(0)}-\frac{\alpha^2 {\ell}^2}{8}\partial^4_y + {\cal O}(\alpha^4)\ .
\ee
 ${\cal L}^{(0)}$ is given (\ref{cmtvlplcin}), which when expressed in terms of  canonical coordinates is
\be\label{L0}
{\cal L}^{(0)}= \left(\ell_0^2+y^2\right)\partial_y^2+(\partial_x+2 y\partial_y)(\partial_x+1)\ .
\ee

Actually, there is a slightly simpler construction.  Instead of mapping ${\cal L}$ to ${\cal L}^{(0)}$, we can map it to ${\cal L}_{\ell}^{(0)}$, which is given by exactly the same expression as in (\ref{L0}), but with $\ell_0 \rightarrow \ell$. For this we just need to set $\ell_1 = 0$ and $\ell_0 = \ell$ in (\ref{Transffinal}), so then $U$ simplifies to (we keep on using the same letter $U$ for the operator)
\bea\label{xpnsn4U}
&U=1+\alpha^2 G+{\cal O}(\alpha^4)\ , \nonumber\\
&G=\frac{1}{96}\Bigl(3 +2 y \partial_y+6\partial_x\Bigr)\partial_y^2+ \frac{3}{32}\frac{1}{\ell^2}( y \partial_y + \partial_x )\ .
\eea
This leads to
\be\label{UcalLUinv}
U{\cal L}U^{-1}={\cal L}_{\ell}^{(0)}+\alpha^2{\cal L}_U^{(1)}+{\cal O}(\alpha^4)
\ee
with
\be\label{LU}
{\cal L}_{\ell}^{(0)}= \left(\ell^2+y^2\right)\partial_y^2+(\partial_x+2 y\partial_y)(\partial_x+1)\ ,\qquad\ {\cal L}_U^{(1)}=-\frac 1 8 {\ell}_0^2\partial^4_y\ .
\ee
The difference between $\ell$ and $\ell_0$ in the second term is of the next order in $\alpha$.

As discussed in Appendix \ref{SectionxyFG}, the quantization in terms of the Fefferman-Graham coordinates is equivalent to making the \textit{commutative} change of variables (\ref{xyFG}) (with $\ell_0$ replaced by $\ell$) in any non-commutative expression written in canonical coordinates. Then transforming  (\ref{LU}) to FG coordinates we get
\be\label{onept4one}
{\cal L}_{\ell}^{(0)}= z^2\left(\partial^2_z + \partial^2_t\right)\ ,\qquad\ {\cal L}^{(1)}_U=-\frac 1{ 8 {\ell}^2}\,z^4\partial^4_t \ .
\ee
So ${\cal L}^{(0)}_\ell$ takes exactly the same form as in (\ref{cmtvlplcin}), but this was only possible only because of the use of the ``quantum'' change of variables (\ref{FGsymbol}). Below we will use the notation ${\cal L}^{(0)}$ instead of ${\cal L}_{\ell}^{(0)}$ assuming that the change $\ell_0 \rightarrow \ell$ is done.

We can now transform the field $\Phi$ to  $\Phi_U=U\Phi$.  From (\ref{fenblk}), it should satisfy the field equation
\be\label{eqfrfiU}
(U{\cal L}U^{-1})\Phi_U= ({m\ell})^2{\Phi_U} \ ,
\ee
which with the help of (\ref{UcalLUinv}) and (\ref{onept4one}) can be written as
\be\label{eqfrfiUxpnd}
\biggl({\cal L}^{(0)}\;-\;\frac {\alpha^2}{ 8 {\ell}^2}\,z^4\partial^4_t \;+\;{\cal O}\Bigl(\frac {\alpha} {\ell}\Bigr)^4 \biggr)\Phi_U(z,t)\;=\;  ({m\ell})^2{\Phi_U}(z,t) \ .
\ee
In order to find the leading  noncommutative corrections to $\Phi^{(0)}$ we can first solve (\ref{eqfrfiU}) for $ \Phi_U$ and then apply the inverse map to get $\Phi$,
\bea\label{phisol}
\Phi(z,t)& =&U^{-1}\Phi_U(z,t) =\biggl(1+ \frac{\alpha^2}{ \ell^2}{\cal D}_{z,t} +{\cal O}\Bigl(\frac{\alpha^4}{\ell^4}\Bigr)\;\biggr)\,\Phi_U(z,t)\ ,
\eea
where
\be\label{calDzt}
{\cal D}_{z,t} :=\frac{z^2}{96}(9+4t\partial_t+6z\partial_z)\partial_t^2+\frac{3}{32}\,z\partial_z \ .
\ee

We note that the leading order correction to $U^{-1}$ vanishes as one approaches the boundary $z\rightarrow 0$,\footnote{Strictly speaking, this is true except for the term $\frac{3}{32}\frac{\alpha^2}{ \ell^2}\,z\partial_z$, which only changes $U^{-1}$ by a constant factor when acting on any power of $z$, but the result  (\ref{fiatbnd}) can also be seen from (\ref{eqfrfiUxpnd}), which in the limit $z\rightarrow 0$ goes to the  commutative equation (\ref{cmtvsffe}).} and so
\be\label{fiatbnd}
\lim_{\epsilon\rightarrow 0}\Phi\Big|_{z=\epsilon}=\lim_{\epsilon\rightarrow 0}\Phi_U\Big|_{z=\epsilon}+{\cal O}\Bigl(\frac{\alpha^4}{\ell^4}\Bigr)\ ,
\ee
while $\partial_z\Phi$ near the boundary gets an additional correction compared  to $\partial_z\Phi_U$:
\be\label{drvfiatbnd}
\lim_{\epsilon\rightarrow 0}\partial_z\Phi|_{z=\epsilon}=\Bigl(1+\frac{3}{32}\frac{\alpha^2}{\ell^2}\Bigr)\,\lim_{\epsilon\rightarrow 0}\partial_z\Phi_U|_{z=\epsilon}\,+\,{\cal O}\Bigl(\frac{\alpha^4}{\ell^4}\Bigr)\ .
\ee

Using standard techniques,\cite{DHoker:2002nbb} one can write down a perturbative solution to (\ref{eqfrfiU}) in terms of a field on the boundary, which we again denote by $\phi_0$. Because noncommutativity vanishes at the boundary (see the further comments in this section and Appendix \ref{AppendixBoundary}), we can assume that  $\phi_0$ is  independent of the perturbation parameter $\alpha/\ell$. Then we  introduce a noncommutative version of the boundary to bulk propagator, denoted by $K^U_{\rm nc}(z,t;t')$, which is defined in analogy to (\ref{frstrdrntsln}),
\be\label{ncfrstrdrntsln}
\Phi_U(z,t)= \int_{ {{R}}} dt'\, K^U_{\rm nc}(z,t;t')\,\phi_0 (t')\ .
\ee
Because we search for perturbative solutions, we expand $\Phi_U$ in even powers of $\alpha/\ell$ about the commutative solution $\Phi^{(0)}$:
\be
\Phi_U=\Phi^{(0)}+\frac{\alpha^2}{{\ell}^2}\Phi^{(1)}+{\cal O}\Bigl(\frac{\alpha^4}{\ell^4}\Bigr)\ .
\ee
From (\ref{eqfrfiU}), $\Phi^{(0)}$ satisfies the free commutative equation (\ref{cmtvsffe}), which is again solved by (\ref{frstrdrntsln}) (with $\ell_0 \rightarrow \ell$), while $\Phi^{(1)}$ satisfies
\be\label{ptbslntldP}
\Bigl({\cal L}^{(0)}- ({m\ell})^2\Bigr)\Phi^{(1)}(z,t)=\frac 1 8 z^4\partial_t^4\Phi^{(0)}(z,t)\ .
\ee
Substituting $\Phi^{(0)}$ from (\ref{frstrdrntsln}) gives
\be\label{EquationPhi1}
\Bigl({\cal L}^{(0)}- ({m\ell})^2\Bigr)\Phi^{(1)}(z,t)=\frac 1 8 z^4\int_{ {{R}}} dt'\, \partial_t^4K(z,t;t')\,\phi_0 (t') \ .
\ee
Next apply the bulk-to-bulk propagator $G(z,t;z',t')$ to obtain an  integral expression  for $\Phi^{(1)}$. Using the conventions in \cite{DHoker:2002nbb}, $G(z,t;z',t')$ satisfies
\bea\label{dfeqfrG}
\biggr\{-(\partial^2_z + \partial_t^2) + \Bigl(\frac{m\ell}z\Bigr)^2\biggl\}\, G(z,t;z',t')&=&\delta(z-z')\delta(t-t')
\eea
and its explicit form in terms of the hypergeometric function is given in Appendix \ref{Asymptotics} (\ref{Bulk2bulk}). Then the solution of (\ref{EquationPhi1}) is given by
\be\label{frstrdrntslncrtn}
\Phi^{(1)}(z,t)= -\frac 18\int_0^\infty dz'z'^2\int_R dt'\,G(z,t;z',t') \int_R dt''\,\partial_{t'}^4 K(z',t';t'') \phi_0 (t'')\ .
\ee
Combining (\ref{frstrdrntsln}) and (\ref{frstrdrntslncrtn}) we obtain an expression for the noncommutative  boundary-to-bulk propagator up to first order in $\alpha^2/\ell^2$:
\be\label{ncbndreblk}
K^U_{\rm nc}(z,t;t')=K(z,t;t')-\frac{\alpha^2}{8\ell^2}\int_0^\infty dz'z'^2\int_R dt''\, G(z,t;z',t'')\, \partial_{t''}^4 K(z',t'';t')+{\cal O}\Bigl(\frac{\alpha^4}{\ell^4}\Bigr)\ .
\ee
From the asymptotic behavior of the commutative Green functions (\ref{AsymptoticsK}) and (\ref{AsymptoticsG}) it follows that $K^U_{\rm nc}(\epsilon,t;t')\;\rightarrow\;K(\epsilon,t;t')$ as $\epsilon\rightarrow 0$, again showing that non-commutative $AdS_2$ is asymptotically \textit{commutative} $AdS_2$.

From the solution (\ref{ncfrstrdrntsln}) for $\Phi_U$, we then get the solution (\ref{phisol}) for $\Phi$. They are functionals of $\phi_0$ and we denote the latter solution by $\Phi_{sol}[\phi_0]$. We next need to substitute $\Phi_{sol}[\phi_0]$ back into (\ref{Sclrfldactn}) to compute the on-shell action. For this purpose it is convenient to re-write the action (\ref{Sclrfldactn}) as
\be
S_{nc}[\Phi]=\frac 1{2\ell\alpha^2}\int_{ {{R}}^2} dxdy\,\biggl\{\Phi\star \Bigl( [{\cal X}^\mu,[{\cal X}_{\mu},\Phi]_\star]_\star +(\alpha\ell m)^2 \Phi\Bigr) -[{\cal X}^\mu,\Phi\star [{\cal X}_{\mu},\Phi]_\star]_\star\biggr\} \ .
\ee
From the field equation, (\ref{fenblk}), the quantity in parenthesis $\Bigl(\cdots\Bigr)$ vanishes on-shell. The  remaining term
\be\label{ncosbndryactn}
S_{nc}^{bdy}[\Phi]= -\frac 1{2\ell\alpha^2} \int dxdy\,[{\cal X}^\mu,\Phi\star [{\cal X}_\mu,\Phi]_\star]_\star \ ,
\ee
is only defined on the boundary ${z=0}$, since the Moyal star-commutator is a total divergence. We argue in the Appendix \ref{AppendixBoundary} that
it contains  no non-commutative corrections, and so, up to an overall factor, has the same form, i.e.,  (\ref{cmtvBndTurm}), as in the commutative theory.
Thus
\bea\label{fortee7}
S_{nc}\Bigl[\Phi_{sol}[\phi_0]\Bigr]&=& S_{nc}^{bdy}\Bigl[\Phi_{sol}[\phi_0]\Bigr]\;=\;-\frac 12\int_{ {{R}}} dt \;\Phi_{sol}[\phi_0] \; \partial_z\Phi_{sol}[\phi_0] \Big|_{z=0}\ .
\eea
Therefore the non-commutative effects are only due the corrections to the solution of the field equation.

Using the near-boundary behavior (\ref{fiatbnd}) and (\ref{drvfiatbnd}), one can express the on-shell action up to order $\frac{\alpha^2}{\ell^2}$ in terms of $\Phi_U$,  and then using (\ref{ncfrstrdrntsln}), it  can be written in terms of  the non-commutative boundary to bulk Green function
\bea\label{Sncphisol}
S_{nc}\Bigl[\Phi_{sol}[\phi_0]\Bigr]&=&-\frac 12\,\Bigl(1+\frac{3}{32}\frac{\alpha^2}{\ell^2}\Bigr)\int_{ {{R}}} dt \;\Bigl( \Phi_U\; \partial_z\Phi_U\Bigr)\Big|_{z=0}+{\cal O}\Bigl(\frac{\alpha^4}{\ell^4}\Bigr)\cr &&\cr
&=&-\frac 12\Bigl(1+\frac{3}{32}\frac{\alpha^2}{\ell^2}\Bigr)\int_R dt \int_R dt'\int_R dt'' K^U_{\rm nc}(z,t;t') \partial_z K^U_{\rm nc}(z,t;t'') \Big|_{z=0} \phi_0 (t')  \phi_0 (t'')\cr&&\cr &&\qquad\qquad\quad+\;{\cal O}\Bigl(\frac{\alpha^4}{\ell^4}\Bigr)\ .
\eea
In the commutative AdS/CFT correspondence it is assumed that the same relation (\ref{adscft}) between the on-shell bulk action and the generating functional for the boundary theory holds for \textit{all} asymptotically $AdS$ spaces. We assume that this continues to be a valid assumption even when the bulk does not correspond to a commutative geometry. Furthermore, after examining the asymptotic behavior of the  Killing vectors, it was argued in \cite{Pinzul:2017wch} that our formulation of noncommutative $AdS^2$ coincides with ``commutative'' $AdS^2$ as one approaches the boundary. So applying (\ref{adscft}) the resulting expression for the two-point correlator on the boundary is
$$ <{\cal O}(t){\cal O}(t')>\,=\,-\frac 12\,\Bigl(1+\frac{3}{32}\frac{\alpha^2}{\ell^2}\Bigr)\int_R dt''\,\Bigl\{ K^U_{\rm nc}(z,t'';t)\, \partial_z K^U_{\rm nc}(z,t'';t') \Big|_{z=0}+\,(t\rightleftharpoons  t')\Bigr\}$$
\be\label{thirte1}
\;+\,\;{\cal O}\Bigl(\frac{\alpha^4}{\ell^4}\Bigr)\ .
\ee
Expanding this in powers of $(\alpha/\ell)^2$:
\be
<{\cal O}(t){\cal O}(t')>=<{\cal O}(t){\cal O}(t ')>^{(0)}+\frac{\alpha^2}{\ell^2}<{\cal O}(t){\cal O}(t')>^{(1)}+\;{\cal O}\Bigl(\frac{\alpha^4}{\ell^4}\Bigr)
\ee
and using (\ref{ncbndreblk}) in (\ref{thirte1}) one recovers (\ref{2ptfncmtv}) at zeroth order, while for the next order we get
$$ <{\cal O}(t){\cal O}(t')>^{(1)}=\,\frac 1{16}\int_0^\infty dz'{z'}^2\int_R dt''\int_R dt'''\,\biggl\{\partial_z\Bigl( K(z,t'';t)\,  G(z,t'';z',t''') \Bigr)\Big|_{z=0}\partial_{t'''}^4  K(z',t''';t')$$
\be
\qquad\qquad\qquad\qquad\qquad+\;\;(t\rightleftharpoons  t') \,\biggr\}\;\;+\;\;\frac{3}{32}<{\cal O}(t){\cal O}(t ')>^{(0)}\ .
\ee
The integrals can be evaluated using the asymptotic obtained in (\ref{AsymptoticsKG}), leading to
\be\label{neyesidntee}
\int dt'\;\partial_{z}\Bigl( K(z,t';t) G(z,t';z',t'')\Bigl)\Big|_{z=0} =\frac 1{2\nu}K(z',t;t'') \ .
\ee
Then the first order correction to the two-point correlation function becomes
\be\label{thurtee5}
<{\cal O}(t){\cal O}(t')>^{(1)}=\frac{1}{32\nu}\;\Bigl({\cal I}_{\Delta_+}(t,t')+{\cal I}_{\Delta_+}(t',t)\Bigr)\;  +\;\frac{3}{32}<{\cal O}(t){\cal O}(t ')>^{(0)}\ ,
\ee
where
\bea\label{ntgrlfrcrtn}
{\cal I}_{\Delta_+}(t,t')&=& \int_0^\infty dz{z}^2\int_R dt''K(z,t;t'') \partial_{t''}^4 K(z,t'';t')\cr &&\cr&=&\partial_{t'}^4\int_0^\infty dz{z}^2\int_R dt''K(z,t;t'')  K(z,t'';t')
\eea
and we used the fact that $K(z,t;t'')$ is only a function of $z$ and $t-t'$.  From the formula (22) of \cite{Freedman:1998tz},
\be
\int_0^\infty dz{z}^2\int_R dt''K(z,t;t'')  K(z,t'';t')=C_{\Delta_+}^2\,\frac{\sqrt{\pi}}{12}\;\frac{ \Gamma(\Delta_++\frac 32) \Gamma(\Delta_+-2)}{\Gamma(\Delta_+)^2 }\;{|t-t'|^{^{4-2\Delta_+}}}\ ,
\ee
which  is analytic for $\Delta_+>2$.  Then using properties of Gamma function,
\bea
{\cal I}_{\Delta_+}(t,t')&=& \frac 43 \,\frac{({\Delta_+}+\frac 12)(\Delta_+-\frac 12)^3(\Delta_+-\frac 32)\,\Gamma(\Delta_+)}{\sqrt{\pi}\,\Gamma(\Delta_++\frac 12)}\;\frac 1{|t-t'|^{^{2\Delta_+}}}\ .
\eea
Substituting into (\ref{thurtee5}) then gives
\be\label{Correlator_free1}
<{\cal O}(t){\cal O}(t')>^{(1)}=\frac {\Gamma(\Delta_+)}{32\sqrt{\pi}\,\Gamma(\Delta_+-\frac 12)}\Big\{\frac 83 \,{\Bigl(\Delta^2_+-\frac 14\Bigr)\Bigl(\Delta_+-\frac 32\Bigl)}{} -{ {{3}\Delta_+}{} }\Bigr\}\frac 1{|t-t'|^{^{2\Delta_+}}} \ .
\ee
Therefore the leading non-commutative correction to the two-point correlator is just a re-scaling of the commutative two-point correlator.

Note that though the zero order correlator, $<{\cal O}(t){\cal O}(t')>^{(0)}$ was calculated for the free commutative theory defined by $m$ and $\ell$, rather than $m_0$ and $\ell_0$, it actually depends on those parameters only through $\Delta_+$, and only in the case when $m_0\ell_0 = m\ell$ does the conformal weight $\Delta_+$  not receive  a leading order non-commutative correction.

For the special case of a massless scalar field, i.e. when $\Delta_+ =1$, (\ref{Correlator_free1}) reduces to
\be
<{\cal O}(t){\cal O}(t')>^{(1)}=-\frac{1}{8\pi}\frac{1}{|t-t'|^2} \ ,
\ee
reproducing the main result of \cite{Pinzul:2017wch}.\footnote{Here we have corrected the error in the numerical factor, which in \cite{Pinzul:2017wch} was erroneously given with an extra factor of $\frac{1}{2}$.}

\section{Interacting scalar field}
\setcounter{equation}{0}
\subsection{Commutative case}

We first review the commutative theory. Upon adding a cubic term to the free scalar field action (\ref{clmsfa}), we get
\bea\label{clmsfawcbctrm}
S[\Phi^{(0)}] = \frac{1}{2}\int_{ {{R}}\times  {{R}}_+} dt dz\,\,\Bigl\{( \partial_z\Phi^{(0)} )^2 \,+\,(\partial_t\Phi^{(0)})^2  \,+\,\Bigl(\frac{m_0\ell_0}z\Bigr)^2{\Phi^{(0)}}^2\;+\; \frac{2\lambda}{3z^2}{\Phi^{(0)}}^3 \Bigr\}\ ,
\eea
where $\lambda$ is a real parameter, and the $(0)$ superscript again indicates that this is the commutative system.
The resulting field equation is now
\be
\Bigl({\cal L}^{(0)}-({m_0\ell_0})^2\Bigr){\Phi^{(0)}}= \lambda\, {\Phi^{(0)}}^2 \ ,  \label{cmlcbce}
\ee
with the same ${\cal L}^{(0)}$ as in (\ref{cmtvlplcin}). We again assume the asymptotic behavior (\ref{smtptccmtvnsr}). Then (\ref{cmlcbce}) can be solved perturbatively in $\lambda$ using the boundary-to-bulk and bulk-to-bulk propagators, $K(z,t;t')$ and $G(z,t;z',t')$, (\ref{Bulk2boundary}) and (\ref{Bulk2bulk}) respectively. Of course, at zeroth order in $\lambda$ the solution is (\ref{frstrdrntsln}). Up to first order one has
\bea
\Phi^{(0)}(z,t)&=&  \int dt'\, K(z,t;t')\,\phi_0 (t') \cr&&\cr &-& {\lambda}\int \frac{dz'dt'}{z'^2} \,G(z,t;z',t')\int  dt_1\int dt_2 \; K(z',t';t_1) K(z',t';t_2)\,\phi_0 (t_1)\phi_0 (t_2)\;\cr&&\cr &+&{\cal O}(\lambda^2)\ .\label{cmtvprtbsln}
\eea
We again denote the solution by $\Phi^{(0)}_{sol}[\phi_0]$.

The on-shell action  now includes a bulk term, as well as a boundary term (which is the same as in (\ref{cmtvBndTurm}))
\bea
S[\Phi^{(0)}]&=&S^{bdy}[\Phi^{(0)}]+S^{blk}[\Phi^{(0)}]\label{cmtvosAktn}\\
&&\cr S^{bdy}[\Phi^{(0)}]&=& -\frac 12\int_{ {{R}}} dt \;\Phi^{(0)}\; \partial_z\Phi^{(0)}\Big|_{z=0}\;,\cr
&&\cr  S^{blk}[\Phi^{(0)}]&=&\frac\lambda 3\int_{ {{R}}\times  {{R}}_+} \frac{dt dz}{z^2}\, {\Phi^{(0)}}^3\ . \label{NCbulk}
\eea
Evaluating the boundary term for $\Phi^{(0)}=\Phi^{(0)}_{sol}[\phi_0]$ gives
\bea\label{NCboundary}
S^{bdy}\Bigl[\Phi^{(0)}_{sol}[\phi_0]\Bigr] &=&-\frac{1}{2} \int dt  dt' dt'' \left[ K(z,t;t')\partial_z K(z,t;t'') \right]_{z=0} \phi_0(t')\phi_0(t'')\nn\\
&&+\frac{\lambda}{2}\int dt \frac{dz' dt'}{z'^2}dt_1 dt_2 dt_3
\partial_z \left( K(z,t;t_1)G(z,t;z',t') \right)\Big|_{z=0}\times \nn \\
&&\times K(z',t';t_2) K(z',t';t_3)\phi_0(t_1)\phi_0(t_2)\phi_0(t_3) + {\cal O}(\lambda^2)  \nn\\
&&  \nn\\
&=& -\frac{\Delta_+ \Gamma (\Delta_+)}{\sqrt{\pi}\Gamma (\Delta_+-\frac{1}{2})}\int dt' dt'' \frac{\phi_0(t')\phi_0(t'')}{|t' - t''|^{2\Delta_+}} \nn\\
&&+\frac{\lambda\Delta_+}{4\nu}\int \frac{dz' dt'}{z'^2}dt_1 dt_2 dt_3 K(z',t';t_1) K(z',t';t_2) K(z',t';t_3)\times \nn\\
&&\times\phi_0(t_1)\phi_0(t_2)\phi_0(t_3) + {\cal O}(\lambda^2)\ ,
\eea
where we used the asymptotic expressions (\ref{AsymptoticsKK}) and (\ref{AsymptoticsKG}). While the first term is exactly (\ref{zrthrdrosa}) and will lead to the same 2-point function (\ref{2ptfncmtv}), the second term will give a non-trivial contribution to the 3-point function. This should be combined with the bulk term (\ref{NCbulk}), which after substitution of $\Phi^{(0)}=\Phi^{(0)}_{sol}[\phi_0]$ takes the form
\bea
S^{blk}\Bigl[\Phi^{(0)}_{sol}[\phi_0]\Bigr] &=& \frac \lambda 3\int \frac{dt dz}{z^2}\int dt_1 dt_2 dt_3\,K(z,t;t_1) K(z,t;t_2) K(z,t;t_3)\,\phi_0(t_1)\phi_0(t_2)\phi_0(t_3)\cr &&\cr &&\qquad +\;{\cal O}(\lambda^2)\ .
\eea
Combining this with (\ref{NCboundary}) and using the definition (\ref{adscft}), the three-point function is
\be\label{cmtatv3ptfn}
<{\cal O}(t_1){\cal O}(t_2){\cal O}(t_3)>^{(0)}=  \lambda \,\Bigl(\frac{3\Delta_+ }{2\nu}+2\Bigr)\int \frac{dzdt}{z^2} \;  K(z,t_1;t)K(z,t;t_2)K(z,t;t_3)\ .
\ee
The dependence on $t_1,\,t_2$ and $t_3$ is determined from conformal invariance
\be\label{zthrdr3pt}
<{\cal O}(t_1){\cal O}(t_2){\cal O}(t_3)>^{(0)}=  \lambda \,\Bigl(\frac{3\Delta_+ }{2\nu}+2\Bigr) \,\frac {a_{\Delta_+}}{|t_1-t_2|^{\Delta_+}|t_2-t_3|^{\Delta_+}|t_3-t_1|^{\Delta_+}}\ .
\ee
The coefficient $ a_{\Delta_+}$ was computed in \cite{Freedman:1998tz}
\be
a_{\Delta_+}=\frac{\Gamma(\Delta_+/2)^3\,\Gamma\Bigl((3\Delta_+-1)/2\Bigr)}{2\pi\,\Gamma(\nu)^3}\ .
\ee

\subsection{Non-commutative case}\label{SectionNCint}

The natural generalization of (\ref{clmsfawcbctrm}) to the noncommutative case is given by
\be\label{cbcactn}
S_{nc}[\hat \Phi]=-\frac 1{2\ell}{\rm Tr}\,\Bigl\{[\hat X^\mu,\hat \Phi][\hat X_\mu,\hat \Phi]-(\alpha\ell m)^2 \hat\Phi^2-\frac 23 \alpha^2\lambda \hat\Phi^3 \Bigr\}\ .
\ee
This action can again be  mapped to an integral on  the  Moyal-Weyl plane
\be\label{Sclrflcbcdactn}
S_{nc}[\Phi]=-\frac 1{2\ell\alpha^2} \int_{ {{R}}^2} dxdy\,\Bigl\{[{\cal X}^\mu,\Phi]_\star\star [{\cal X}_{\mu},\Phi]_\star-(\alpha\ell m)^2 \Phi\star\Phi  -\frac 23 \alpha^2\lambda\, \Phi\star\Phi\star\Phi\Bigr\}\ .
\ee
The field equation  following from (\ref{Sclrflcbcdactn}) is
\bea\label{fewficub}
{\cal L}\Phi -(\ell m)^2 \Phi&=&\lambda \,\Phi\star\Phi \ .
\eea
where $ {\cal L}$ is the noncommutative Laplace operator, defined by (\ref{fenblk}), (\ref{nclplnxy}) or (\ref{Klngcmtvsffenc}).

The free theory, $\lambda=0$, is solved by (\ref{phisol}), (\ref{ncfrstrdrntsln}) and (\ref{ncbndreblk})
\bea
&&\Phi(z,t)\;=\; U_{z,t}^{-1}\Phi_U(z,t)\;=\;  \int dt'\, K_{\rm nc}(z,t;t')\,\phi_0 (t')\ ,
\eea
where
\bea\label{ncbndblkgf}
K_{\rm nc}(z,t;t')&=& U_{z,t}^{-1} K^U_{\rm nc}(z,t;t')
\cr&&\cr
&=&K^U_{\rm nc}(z,t;t')+\frac{\alpha ^2} { \ell_0^2}{\cal D}_{z,t}K(z,t;t')+{\cal O}\left(\alpha ^4\right)\ ,
\eea
with the differential operator $ {\cal D}_{z,t}$ defined in (\ref{calDzt}).

For  small $\lambda$, (\ref{fewficub}) can be solved  perturbatively  in analogy with (\ref{cmtvprtbsln}), by  replacing  the commutative  source $\lambda{\Phi^{(0)}}^2$ by its non-commutative analogue $\lambda \,\Phi\star\Phi$, and by replacing the commutative Green functions by their non-commutative analogues. The non-commutative analogue of the  boundary to bulk Green function is given by (\ref{ncbndblkgf}). We denote the non-commutative analogue of the bulk to bulk Green function by $G^U_{\rm nc}(z,t;z',t')$. We require it to satisfy the analog of the commutative equation (\ref{dfeqfrG})
\bea
\Bigl[(U {\cal L}U^{-1})_{z,t} -(\ell m)^2 \Bigr]G^U_{\rm nc}(z,t;z',t')&=&-z^2\,\delta(z-z')\delta(t-t')\ ,
\eea
so $G^U_{\rm nc}(z,t;z',t')\rightarrow G(z,t;z',t')$ in the commutative limit. Upon expanding in $\lambda$, the solution to (\ref{fewficub}) is
\bea\label{nccmtvprtbsln}
\Phi(z,t) &=&  \int dt'\, K_{\rm nc}(z,t;t')\,\phi_0 (t')\qquad\qquad\qquad\cr&&\cr
&-&{\lambda}\int \frac{dz'dt' }{z'^2}\,U^{-1}_{z,t} G^U_{\rm nc}(z,t;z',t')\int dt_1 dt_2  \,U_{z',t'}[  K_{\rm nc}^{(1)}\star K_{\rm nc}^{(2)}](z',t') \;\phi_0 (t_1)\phi_0 (t_2)\cr&&\cr &+&{\cal O}(\lambda^2)\ ,
\eea
where $K_{\rm nc}^{(n)}(z,t)$ denotes the function $ K_{\rm nc}(z,t;t_n)$ and the star-product is with respect to the explicitly shown variables. The solution to $G^U_{\rm nc}(z,t;z',t')$ can be computed perturbatively in powers of $\alpha^2$. If we write
\be\label{NCGU}
G^U_{\rm nc}(z,t;z',t')= G(z,t;z',t')+\alpha^2 G^{(1)}(z,t;z',t')+{\cal O}(\alpha^4)\ ,
\ee
then the leading order non-commutative correction $ G^{(1)}(z,t;z',t')$ satisfies
\bea
\Bigl[{\cal L}^{(0)}_{z,t}-(\ell m)^2 \Bigr]\, G^{(1)}(z,t;z',t')&=& \frac 1{ 8 {\ell}_0^2}\,z^4\partial^4_t  \,G(z,t;z',t')\ ,
\eea
where we used (\ref{eqfrfiUxpnd}). The solution is
\be
G^{(1)}(z,t;z',t')=-\frac 1{ 8 {\ell}_0^2}\int {dz''dt''}{{z''}^2}G(z,t;z'',t'')\,\partial^4_{t''}  \,G(z'',t'';z',t') \ .
\ee

We denote the solution (\ref{nccmtvprtbsln}) by $\Phi_{sol}[\phi_0]$ and substitute it back in (\ref{Sclrflcbcdactn}) to get the on-shell action. The latter can be split into two terms as in the commutative case
\bea
S_{nc}[\Phi]&=&S^{bdy}_{nc}[\Phi]+S^{blk}_{nc}[\Phi] \ ,
\eea
where $S_{nc}^{bdy}[\Phi]$ was defined in (\ref{ncosbndryactn}) and
\be
S_{nc}^{blk}[\Phi]=\frac\lambda {3\ell}\int_{R^2} dxdy \, \Phi\star\Phi\star\Phi \ .
\ee
Substituting  $\Phi_{sol}[\phi_0]$ into the bulk term $S_{nc}^{blk}[\Phi]$ and converting to Fefferman-Graham coordinates (\ref{FGsymbol}) gives
\be\label{EigtheeOne}
\frac \lambda 3\int \frac{dzdt}{z^2}\,dt_1\,dt_2\,dt_3\;[  K_{\rm nc}^{(1)}\star  K_{\rm nc}^{(2)}\star K_{\rm nc}^{(3)}](z,t)\,\phi_0 (t_1)\phi_0 (t_2)\phi_0 (t_3)\ .
\ee
Substituting  $\Phi_{sol}[\phi_0]$ into the boundary term  $S_{nc}^{bdy}[\Phi]$, which again reduces to (\ref{fortee7}), and collecting the third order terms in $\phi_0$, we get
\bea\label{Eigthee}
&& \frac \lambda 2 \int \frac{dz'dt'}{z'^2}dt\,dt_1\,dt_2\,dt_3\;\partial_z\Bigl( K_{nc}(z,t;t_1) U^{-1}_{z,t} G^U_{nc}(z,t;z',t')\Bigl)\Big|_{z=0}
\cr&&\cr&&\qquad\;\;\qquad\;\;\qquad\;\;\times\; \; \,U_{z',t'} [  K_{\rm nc}^{(2)}\star K_{\rm nc}^{(3)}](z',t')\,\phi_0 (t_1)\phi_0 (t_2)\phi_0 (t_3)\ .
\eea
In the commutative limit, we recover the  commutative boundary term (\ref{NCboundary}). This follows from $ K_{\rm nc}(z,t;t')\rightarrow  K(z,t;t')$, $G^U_{\rm nc}(z,t;z',t')\rightarrow  G(z,t;z',t')$ and $U_{z,t} \rightarrow 1$, as $\alpha\rightarrow0$ along with the identity (\ref{AsymptoticsKG}).

The sum of (\ref{Eigthee}) and (\ref{EigtheeOne}) gives all the $\phi_0^3$ terms in $ S_{nc}\Bigl[\Phi_{sol}[\phi_0]\Bigr]$. So the expression for the three-point function is
\bea\label{thuhnc3ptfn}
&&<{\cal O}(t_1){\cal O}(t_2){\cal O}(t_3)>\;=\;\frac{\delta^3 S_{nc}\Bigl[\Phi_{sol}[\phi_0]\Bigr]}{\delta \phi_0(t_1)\delta \phi_0(t_2)\delta \phi_0(t_3)}\bigg|_{\phi_0=0} = \cr&&\cr&&\cr
&=& \frac{\lambda } {2}\;\int \frac{dz\, dt}{z^2}\biggl\{\,\int dt'\;\partial_{z'}\Bigl( K_{\rm nc}(z',t';t_1) U^{-1}_{z',t'}G^U_{\rm nc}(z',t';z,t)\Bigl)\Big|_{z'=0}\cdot\, U_{z,t}[  K_{\rm nc}^{(2)}\star K_{\rm nc}^{(3)}](z,t)
\cr&&\cr& &\qquad\qquad\qquad+ \;\frac 23\;[  K_{\rm nc}^{(1)}\star  K_{\rm nc}^{(2)}\star K_{\rm nc}^{(3)}](z,t)\,\biggr\} \,+\,{\rm all}\;{\rm permutations}\;{\rm of} \;(t_1,t_2,t_3)\ .\cr&&
\eea
Near the boundary the first term in the integrand can be expanded in $\alpha^2$ using the results of the appendix \ref{Asymptotics}. Using the relevant asymptotics in the definitions (\ref{ncbndreblk}), (\ref{ncbndblkgf}) and (\ref{NCGU}) one easily establishes the following asymptotic formulas:
\bea
K_{\rm nc}(z,t;t')&\rightarrow&  z^{1-\Delta_+}\Bigl(1+\frac{3}{32}\frac{\alpha ^2}{\ell^2}(1-\Delta_+)\Bigr)\delta(t-t')\;+\;{\cal O}\left(\alpha ^4\right) \ , \cr&&\cr
\partial_z K_{\rm nc}(z,t;t')&\rightarrow&  z^{-\Delta_+}\Bigl(1-\Delta_+ +\frac{3}{32}\frac{\alpha ^2}{\ell^2}(1-\Delta_+)^2 \Bigr)\delta(t-t')\;+\;{\cal O}\left(\alpha ^4\right) \ , \cr&&\cr
U^{-1}_{z,t}G^U_{\rm nc}(z,t;z',t')&\rightarrow& \frac{1}{2\Delta_+ -1}\, z^{\Delta_+}K^U_{\rm nc}(z',t';t)\Bigr( 1+\frac{3}{32}\frac{\alpha ^2}{\ell^2}\Delta_+  \Bigr) +{\cal O}({\alpha^4})\ ,\cr&&\cr
\partial_z (U^{-1}_{z,t}G^U_{\rm nc}(z,t;z',t'))&\rightarrow& \frac{\Delta_+}{2\Delta_+ -1}\, z^{\Delta_+ - 1}K^U_{\rm nc}(z',t';t)\Bigr( 1+\frac{3}{32}\frac{\alpha ^2}{\ell^2}\Delta_+  \Bigr) +{\cal O}({\alpha^4})\ ,
\eea
which leads to the $z\rightarrow +0$ value for the relevant term in (\ref{thuhnc3ptfn})
\be\label{fnlnc3ptfn}
\partial_z\Bigl(K_{\rm nc}(z,t;t_1)U^{-1}_{z,t}G^U_{\rm nc}(z,t;z',t')\Bigr)\Big|_{z=0}\rightarrow \frac{1}{2\Delta_+ -1} \biggl(1+ \frac{3}{32}\frac{\alpha ^2}{\ell^2} \biggr)K^U_{\rm nc}(z',t';t)\delta(t-t_1) +{\cal O}({\alpha^4})\ .
\ee
Substituting into (\ref{thuhnc3ptfn}) gives
\bea\label{nc3ptcf}
&&<{\cal O}(t_1){\cal O}(t_2){\cal O}(t_3)> \cr &&\cr
&&=\;\frac { \lambda}{2}\;\int \frac{dz\, dt}{z^2}\biggl\{\frac{1}{2\Delta_+ -1}\Bigl(1+\frac{3}{32}\frac{\alpha ^2}{\ell^2}  \Bigr)U_{z,t}K_{\rm nc}^{(1)}(z,t)\cdot\, U_{z,t}[  K_{\rm nc}^{(2)}\star K_{\rm nc}^{(3)}](z,t)
\cr&&\cr& &+ \;\frac 23\;[  K_{\rm nc}^{(1)}\star  K_{\rm nc}^{(2)}\star K_{\rm nc}^{(3)}](z,t)\,\biggr\} \,+\,{\rm all}\;{\rm permutations}\;{\rm of} \;(t_1,t_2,t_3)\cr&&\cr &&+\;{\cal O}\left(\alpha ^4\right)\ .
\eea

Next we will analyze the result (\ref{nc3ptcf}) and demonstrate that it has the same scaling and translational transformation properties as the commutative $3-$point function (\ref{zthrdr3pt}) (at least up to leading order in $\alpha^2$).\footnote{Typically, in order to determine whether a $3-$point function has the form  (\ref{zthrdr3pt}) (up to an overall factor),  one would also have to check its behaviour under  special conformal transformations.  However,  for the case of a 3-point function for the \textit{same} field, or fields with the same conformal dimension, it is sufficient to ensure that the correlator behaves correctly under scaling and translations. It would be a  nice check to demonstrate that (\ref{nc3ptcf}) does indeed  transform as  (\ref{zthrdr3pt})  under special conformal transformations, but this task appears to be quite nontrivial.} First, we will establish the behaviour of (\ref{nc3ptcf}) under the simultaneous scaling of $t_i$, $i=1,2,3$: $t_i\rightarrow \mu t_i$, where $\mu$ is a constant parameter. Using (\ref{Bulk2boundary}), (\ref{Bulk2bulk}), (\ref{calDzt}), (\ref{ncbndreblk}) and (\ref{ncbndblkgf}) one can easily see that under the simultaneous rescalling of \textit{all} the variables the relevant quantities have the following behaviour:
\be
K(\mu z,\mu t; \mu t')=\mu^{-\Delta_+} K(z,t; t')\ ,\;\quad G(\mu z,\mu t;\mu z',\mu t')=G(z,t;z',t')\ ,\quad\; U^{-1}_{\mu z,\mu t}=U^{-1}_{z,t}\nn
\ee
and then
\be\label{Scalings}
K^U_{\rm nc}(\mu z,\mu t;\mu t')=\mu^{-\Delta_+} K^U_{\rm nc}(z,t;t')\ ,\qquad K_{\rm nc}(\mu z,\mu t;\mu t')=\mu^{-\Delta_+} K_{\rm nc}(z,t;t')\ ,
\ee
at least up to order $\alpha^2$. Noticing that the star-product (\ref{dffstrprd}) is constructed from the derivatives $\frac\partial{\partial x}= -z\frac\partial{\partial z}-t\frac\partial{\partial t}$ and $\frac\partial{\partial y}=\frac z{\ell} \frac\partial{\partial t}$, which are invariant under a simultaneous re-scaling of $z$ and $t$, we see that the star-product is scale invariant to all orders. Using this, along with (\ref{Scalings}) in (\ref{nc3ptcf}), we find that the non-commutative $3-$point function scales just like the commutative $3-$point function (\ref{zthrdr3pt})
\be\label{toohndrd}
<{\cal O}(\mu t_1){\cal O}(\mu t_2){\cal O}(\mu t_3)>=\mu^{-3\Delta_+}\,<{\cal O}(t_1){\cal O}(t_2){\cal O}(t_3)>\ .
\ee

Now we would like to demonstrate the invariance of (\ref{nc3ptcf}) under simultaneous translations $t_i\rightarrow t_i + a$. This will effectively guarantee that the non-commutative 3-point function has the same functional dependence on $t_i$ as its commutative counterpart (\ref{zthrdr3pt}). Because both commutative propagators, (\ref{Bulk2boundary}) and (\ref{Bulk2bulk}), depend only on the time translational invariant combination, they are time translationally invariant. Using this in (\ref{ncbndreblk}), we see that this is also true for $K^U_{\rm nc}(z,t;t')$ (at least up to order $\alpha^2$). The first non-trivial effect due to translations appears in the transformation of $U^{-1}(\mu z,\mu t)$. Under $t\rightarrow t+a$ it transforms as
\begin{equation}\label{Utrans}
U^{-1}_{z,t+a}= U^{-1}_{z,t} + \frac{a\alpha^{2} z^{2}}{24\ell}\partial_{t}^{3} + {\cal O}\left(\alpha ^4\right) =: U^{-1}_{z,t} + \mathcal{J}_{z,t}(a) + {\cal O}\left(\alpha ^4\right)\ ,
\end{equation}
where
\begin{equation}\label{Jzt}
\mathcal{J}_{z,t}(a)= \frac{a\alpha^{2} z^{2}}{24\ell^{2}}\partial_{t}^{3}\ .
\end{equation}
This leads to a non-trivial transformation of $K_{\rm nc}(z,t;t')$:
\be
K_{\rm nc}(z,t+a;t'+a)= K_{\rm nc}(z,t;t') + \mathcal{J}_{z,t}(a)K(z,t;t')+ {\cal O}\left(\alpha ^4\right)\ .
\ee
The second non-trivial contribution comes from the star-product. Expanding (\ref{dffstrprd}) up to the second order in $\alpha$ gives
\bea\label{starexpended}
[ {\cal F}\star  {\cal G} ](z,t) &=&  {\cal F}{\cal G} -\frac{i\alpha}{2}z^2\left({\partial_z}{\cal F}{\partial_{t}}{\cal G} - {\partial_{t}}{\cal F}{\partial_z}{\cal G} \right) + \nn\\
&&+\frac{\alpha^{2}z^{2}}{8\ell^{2}}\Big( z^{2}(2\partial_{z}\partial_{t}{\cal F}\partial_{z}\partial_{t}{\cal G} - \partial_{z}^{2}{\cal F}\partial_{t}^{2}{\cal G}-\partial_{z}^{2}{\cal G}\partial_{t}^{2}{\cal F}) + 2\partial_{t}{\cal F}\partial_{t}{\cal G} \nn\\
&&+z(2\partial_{z}\partial_{t}{\cal F}\partial_{t}{\cal G}+2\partial_{z}\partial_{t}{\cal G}\partial_{t}{\cal F}-\partial_{z}{\cal F}\partial_{t}^{2}{\cal G}-\partial_{z}{\cal G}\partial_{t}^{2}{\cal F})\nn\\
&&+t(\partial_{t}^{2}{\cal F}\partial_{t}{\cal G}+\partial_{t}^{2}{\cal G}\partial_{t}{\cal F})\Big)+ {\cal O}\left(\alpha ^3\right)\ .
\eea
Then up to order $\alpha^{2}$ the translated star product gets an extra term:
\begin{equation}
\star_{z,t+a} = \star_{z,t}+\mathcal{S}_{z,t}(a)+ {\cal O}\left(\alpha ^3\right)\ ,
\end{equation}
where
\begin{equation}\label{Szt}
\mathcal{S}_{z,t}(a) := \frac{a\alpha^{2}z^{2}}{8\ell^{2}}\Big(\overleftarrow{\partial}_{t}^2\overrightarrow{\partial}_{t}+\overleftarrow{\partial}_{t}\overrightarrow{\partial}_{t}^2\Big)\ .
\end{equation}

Using the definitions (\ref{Jzt}) and (\ref{Szt}), one can easily verify the following useful relation between $\mathcal{J}_{z,t}(a)$ and $\mathcal{S}_{z,t}(a)$
\bea\label{JKK}
&&\mathcal{J}_{z,t}(a)\Big({\cal F}(z,t){\cal G}(z,t)\Big) \equiv \frac{\alpha^{2}z^{2}a}{24\ell^{2}}\partial^{3}_{t}\Big({\cal F}(z,t){\cal G}(z,t)\Big) = \nn\\
&&={\cal F}(z,t)\mathcal{S}_{z,t}(a){\cal G}(z,t) + {\cal F}(z,t)\mathcal{J}_{z,t}(a){\cal G}(z,t)+{\cal G}(z,t)\mathcal{J}_{z,t}(a){\cal F}(z,t)\
\eea
for any two arbitrary functions ${\cal F}(z,t)$ and ${\cal G}(z,t)$. Using this relation we get ($\mathcal{J}_{z,t}\equiv \mathcal{J}_{z,t}(a)$)
\bea\label{2nd1}
&&{K}_{nc}(z,t+a;t_{2}+a)\star_{z,t+a}{K}_{nc}(z,t+a;t_{3}+a)=\nn\\
&&={K}_{nc}(z,t;t_2)\star_{z,t}{K}_{nc}(z,t;t_3) + \mathcal{J}_{z,t}\Big(K(z,t;t_2)K(z,t;t_3) \Big)+ {\cal O}\left(\alpha ^3\right)
\eea
and
\bea\label{KKK}
&&{K}_{nc}(z,t;t_1+a)\star_{z,t+a}{K}_{nc}(z,t;t_2+a)\star_{z,t+a}{K}_{nc}(z,t;t_3+a) = \nn\\
&&={K}_{nc}(z,t;t_1)\star_{z,t}{K}_{nc}(z,t;t_2)\star_{z,t}{K}_{nc}(z,t;t_3)+\nonumber \\
&&+\mathcal{J}_{z,t}\Big( K(z,t;t_1)K(z,t;t_2)K(z,t;t_3)\Big) + {\cal O}\left(\alpha ^3\right)\ ,
\eea
where in non-invariant terms we used ${K}(z,t;t_i)$ instead ${K}_{nc}(z,t;t_i)$ because these terms are already of the order of $\alpha^2$. Combining (\ref{Utrans}) and (\ref{2nd1}) and using the translational invariance of $K^U_{\rm nc}(z,t;t')$, we see that the contribution of the boundary term to the 3-point function (\ref{nc3ptcf}) is explicitly translationally invariant. The bulk contribution to the 3-point function is also translationally invariant due to the fact that the non-invariant term coming from (\ref{KKK}) is given by integral of a total derivative, which will drop out of (\ref{nc3ptcf}) even without symmetrization.

We have demonstrated that the non-commutative 3-point function (\ref{nc3ptcf}) shares the symmetries of its commutative counterpart and, as a consequence, should have the following form
\be\label{nc3pt}
<{\cal O}(t_1){\cal O}(t_2){\cal O}(t_3)> = \left(1 + c\alpha^2 \right) <{\cal O}(t_1){\cal O}(t_2){\cal O}(t_3)>^{(0)}+ {\cal O}\left(\alpha ^3\right)\ ,
\ee
where $<{\cal O}(t_1){\cal O}(t_2){\cal O}(t_3)>^{(0)}$ is given in (\ref{zthrdr3pt}) and the coefficient $c$ should be calculated by an explicit evaluation of the terms in (\ref{nc3ptcf}) up to the order $\alpha^2$.

\section{Conclusion}

In this paper we have tested the possibility of extending the $AdS/CFT$ correspondence to the case of a non-commutative bulk. Making the domain on the gravity-side of the correspondence  non-commutative is physically well motivated for, at least, three reasons:
\begin{enumerate}
\item  As  was mentioned in the introduction,  non-commutative space-time can be interpreted as a quasi-classical regime of essentially any theory of quantum gravity. Because the strong form of the $AdS/CFT$ correspondence assumes a duality between full quantum gravity (and not just classical gravity) and $CFT$ living on the conformal boundary, it is reasonable to assume that the duality can be extended to the quasi-classical regime as well.
\item It is believed that the correspondence should work for any asymptotically $AdS$ space. As we demonstrated in \cite{Pinzul:2017wch}, $ncAdS$ has this property - the non-commutativity effectively vanishes near the boundary.
\item The deformation leading to $ncAdS$ used in this paper preserves the \textit{undeformed} $SO(2,1)$ isometry of the bulk, which supports the notion that a dual theory will have  \textit{undeformed}  $SO(2,1)$ conformal symmetry.
\end{enumerate}

To test whether the $AdS/CFT$ correspondence can be extended to a non-commutative setting, we have calculated the leading non-commutative corrections to 2- and 3-point correlation functions for a scalar field on the non-commutative background. To get a non-vanishing 3-point function, the field has to be interacting. This created serious technical complications compared to the free case. Still we were able to show that for both correlators the overall effect of non-commutativity is in re-scaling of their commutative counterparts, supporting and generalizing the conclusions of \cite{Pinzul:2017wch} for the free massless case. For the 2-point function we were able to compute the answer in a closed form (\ref{Correlator_free1}), explicitly demonstrating its conformal behaviour (and greatly simplifying the analysis of \cite{Pinzul:2017wch}).  On the other hand, due to the aforementioned technical complications, we were not able to obtain a corresponding closed form expression for the 3-point function. There instead  the analysis was done implicitly by studying the transformation properties of the correlator under the conformal transformations, which also confirmed its conformal behaviour.

This result prompts several immediate questions. The first one is about the triviality of the result. Is  it possible that the commutative theory can be mapped to the non-commutative one, analogous to the Seiberg-Witten map for gauge theories?  More explicitly, can the non-commutative Killing vectors be obtained from the commutative ones by some similarity transformation? If true, this would give a trivial solution to the non-commutative field equations. In this paper we gave a perturbative argument to show that this is not the case, and so the field equation (and as the result, the solution) receive a genuinely non-commutative correction, proving that our result is really non-trivial.

The second question one can ask is whether or not the whole effect of the non-commutativity (associated with a quantum gravitational bulk) on the  n-point functions of a dual $CFT$ is due to some kind of non-commutative (or quantum gravitational) renormalization of the corresponding boundary operator $\mathcal{O}$. To address this issue, one has to compare the factors in front of the non-commutative 2- and 3-point correlators. The question then is are the corrections to the correlators related by a factor of $2/3$? While we have an  explicit form (\ref{Correlator_free1}) for the leading correction to the 2-point correlator, due to technical complications we were unable to do the integration in  (\ref{nc3pt}) to get  the explicit form for the 3-point  correlator. This is a very important technical problem, whose solution would give a better understanding of the correspondence. We are planning to report on the progress in this direction elsewhere.

An obvious question is whether or not our perturbative analysis can be generalized beyond the first non-trivial order, and furthermore if the conclusion we found at leading order survives? What gives us hope that  the conclusion does indeed survive is  the presence of the full {\it  undeformed} symmetry in the non-commutative theory, which might lead to the construction of some exact non-perturbative results. It is important to have this analysis done to ensure that our findings are not an artifact of the first order approximation. We are planning to look at this problem as one of the natural continuations of the project.

The final question concerns the possibility of  generalizing of the noncommutative construction presented in this paper to higher dimensions.  This is particularly relevant because the (commutative) $AdS_{d+1}/CFT_d$ correspondence is best understood for $d\ge 2$.  It should be possible to apply the arguments presented  in this paper  to  a non-commutative  $AdS_{d+1}/CFT_d, \;d\ge 2$  correspondence provided that one first succeeds to construct an isometry preserving non-commutative deformation of $AdS_{d+1}$. The problem, however, is that in more then two dimensions, due to the lack of a natural Poisson structure, this is a much more difficult, both technically and conceptually, task.  One possible exception is for  $AdS_4$.\cite{Sperling:2018xrm} We hope to report some preliminary results on this case of in the near future.

\appendix
\setcounter{equation}{0}
\section{Useful asymptotic expressions}\label{Asymptotics}

Here we calculate some asymptotics of the Green's functions used in the main text. The definitions of the boundary-to-bulk and bulk-to-bulk propagators, respectively, are as follows:
\bea
K(z,t;t'):=K_{\Delta}(z,t;t')=C_{\Delta}\left( \frac{z}{z^2 + (t-t')^2} \right)^\Delta \ ,\quad C_{\Delta}=\frac{\Gamma (\Delta)}{\sqrt{\pi}\Gamma (\Delta -\frac{1}{2})}\ ,\label{Bulk2boundary} \\
G(\xi):= \frac{C_{\Delta}}{2\Delta -1}\left( \frac{\xi}{2} \right)^{\Delta} F\left(\frac{\Delta}{2},\frac{\Delta}{2}+\frac{1}{2};\Delta+\frac{1}{2};\xi^2\right)\ ,\qquad \xi=\frac{2zz'}{z^2+z'^2+(t-t')^2} \label{Bulk2bulk}
\eea
and we defined $\Delta_+ = \Delta$, $\Delta_- =1-\Delta$, while $F(a,b;c;z)\equiv {}_2 F_1 (a,b;c;z)$ is the standard hypergeometric function normalized as $F(a,b;c;0)=1$. Taking the derivative of (\ref{Bulk2boundary}) with respect to $z$, one finds
\bea\label{derK}
\partial_z K_{\Delta}(z,t;t')=\frac{\Delta}{z} K_{\Delta}(z,t;t') -(2\Delta -1) K_{\Delta +1}(z,t;t') \nn\ .
\eea
Combining this with the asymptotics for $K(z,t;t')$, which trivially follows from (\ref{smtptccmtvnsr}) and (\ref{frstrdrntsln}) one gets
\bea\label{AsymptoticsK}
K(z,t;t') &\xlongrightarrow[z\rightarrow 0]{}& z^{1-\Delta} \delta(t-t')\ , \nn\\
\partial_z K(z,t;t') &\xlongrightarrow[z\rightarrow 0]{}& (1-\Delta)z^{-\Delta} \delta(t-t')\ .
\eea

One should be careful in taking the $z\rightarrow 0$ limit for $K(z,t;t')\partial_z K(z,t;t'')$. First assume that $|t'-t''|\gg \epsilon >0$. From (\ref{Bulk2boundary}) one gets
\bea
K(z,t;t')\partial_z K(z,t;t'') = z^{\Delta - 1}K(z,t;t')\Delta C_{\Delta}\left( \frac{1}{z^2 + (t-t'')^2} \right)^{\Delta}\,\frac{-z^2 + (t-t'')^2}{z^2 + (t-t'')^2}\ .
\eea
Using (\ref{AsymptoticsK}) one more time and taking into account that $|t'-t''|\gg \epsilon >0$ we obtain
\bea\label{AsymptoticsKK}
K(z,t;t')\partial_z K(z,t;t'') &\xlongrightarrow[z\rightarrow 0]{}& \Delta C_{\Delta}\delta(t-t')\frac{1}{|t'-t''|^{2\Delta}}\ ,
\eea
which is now valid for any $t'\ne t''$ on the boundary.

To get the analogous results for the bulk-to-bulk propagator we just use the expansion of the hypergeometric function in the definition (\ref{Bulk2bulk})
\bea
G(\xi)= \frac{C_{\Delta}}{2\Delta -1}\left( \frac{\xi}{2} \right)^{\Delta} \left(1+ \frac{\frac{\Delta}{2}\left(\frac{\Delta}{2}+\frac{1}{2}\right)}{\Delta+\frac{1}{2}}\xi^2 + \mathcal{O}(\xi^4)\right)\ .\nn
\eea
Then taking into account
\bea
\xi=\frac{2zz'}{z'^2+(t-t')^2} + \mathcal{O}(z^3)\quad \text{and}\quad \partial_z \xi = \frac{1}{z}\xi - \frac{1}{z'}\xi^2
\eea
we immediately get
\bea\label{AsymptoticsG}
G(\xi) &\xlongrightarrow[z\rightarrow 0]{}& \frac{C_{\Delta}}{2\Delta -1}\left( \frac{z'}{z'^2+(t-t')^2} \right)^{\Delta}z^{\Delta}\equiv \frac{1}{2\Delta -1}z^{\Delta}K(z',t';t) \ , \nn\\
\partial_z G(\xi) &\xlongrightarrow[z\rightarrow 0]{}& \frac{\Delta}{2\Delta -1}z^{\Delta -1}K(z',t';t)\ .
\eea

We also need to evaluate $z\rightarrow 0$ behaviour of $\partial_z \left( K(z,t;t')G(z,t;z'',t'') \right)$ which is now trivially found using (\ref{AsymptoticsK}) and (\ref{AsymptoticsG})
\bea\label{AsymptoticsKG}
\partial_z \left( K(z,t;t')G(z,t;z'',t'') \right) \xlongrightarrow[z\rightarrow 0]{} \frac{1}{2\Delta -1}\delta(t-t')K(z'',t'';t')\ .
\eea

\section{Noncommutative canonical coordinates versus noncommutative Fefferman-Graham coordinates}\label{SectionxyFG}
\setcounter{equation}{0}

As  was discussed in detail in \cite{Pinzul:2017wch}, the canonical coordinates (\ref{corrd_canon}) have the canonical Poisson structure (hence the name). Via (\ref{xyFG}) this induces the following Poisson brackets for the FG coordinates appearing in (\ref{corrd_FG})
\be\label{FGPoisson}
\{ t,z \}=\frac{1}{\ell_0}z^2 \ .
\ee
While the quantization of the canonical structure leads to the Moyal-Weyl star product (\ref{dffstrprd}), the quantization of (\ref{FGPoisson}) is not that obvious due to the non-triviality of the Poisson structure in terms of FG coordinates. Here we argue that the most natural choice of the quantization, the symmetric one, leads to the minimal quantization of (\ref{FGPoisson}), i.e. does not introduce the higher-order in $\alpha$ terms.

Towards this end, we upgrade $t$ and $z$ to operators using the definitions
\be\label{FGNC}
\begin{array}{l}
  t= \frac{1}{2\ell}(y e^{-x}+e^{-x}y)\\
  z= e^{-x}
\end{array}
\ ,
\ee
where here $x$ and $y$ are also interpreted as operators.
Using the canonical commutation relation (\ref{canonic_comm}) for $x$ and $y$, one readily finds the commutator between $t$ and $z$
\be\label{comm_FG}
[t,z]=i\frac{\alpha}{\ell}z^2 \ .
\ee
So, this is really the minimal quantization of (\ref{FGPoisson}). Re-written in terms of the star-product (\ref{dffstrprd}), it will become the star-commutator between the corresponding symbols, which are, as trivially verified, given by the commutative expression (\ref{xyFG}) (without any star-product!) but now with $\ell_0 \rightarrow \ell$
\be\label{FGsymbol}
\begin{array}{l}
  t= \frac{1}{\ell}y e^{-x}\\
  z= e^{-x}
\end{array}
\ .
\ee
Now we are regarding $x$, $y$, $t$ and $z$ as symbols of the corresponding operators. Then the star-product in terms of the Fefferman-Graham coordinates will be given just by the standard Moyal-Weyl product (\ref{dffstrprd}) in which one performs the commutative change of variables (\ref{FGsymbol}). One could also explicitly verify this by calculating the star-product and observing that $z\star z=e^{-x}\star e^{-x}\equiv z^2$ and $\frac{1}{2}(y\star e^{-x} + e^{-x}\star y)=y e^{-x}$.

\section{Similarity transformation and no-triviality}\label{Similar}
\setcounter{equation}{0}

Both sets of Killing vectors, commutative and noncommutative, satisfy \textit{the same} undeformed $so(2,1)$ algebra: $[K^\mu, K^\nu]=\epsilon^{\mu\nu\rho}K_\rho$. The expressions for the Killing vectors  in terms of the canonical coordinates was given in \cite{Pinzul:2017wch}. The commutative Killing vectors are
\begin{eqnarray}\label{Killings}
\left\{
\begin{array}{l}
  K^0=\partial_x \\
  K^-=  -\ell_0 e^{x}\,\partial_y \\
  K^+=  \frac 1{\ell_0} e^{-x}\left(2 y\,\partial_x\,+(y^2 + \ell_0^2)\partial_y \right)
\end{array}
\right. \ ,
\end{eqnarray}
while the non-commutative Killing vectors are
\begin{eqnarray}\label{ncKillings}
\left\{
\begin{array}{l}
  K_\star^0=\partial_x\equiv K^0 \\
  K_\star^-=  -\ell e^{x}\,\Delta_y \\
  K_\star^+=  \frac 1{\ell} e^{-x}\left(2 y\,\partial_x S_y\,+(y^2 + \ell^2 + \frac{\alpha^2}{4}(1-\partial_x^2))\Delta_y \right)
\end{array}
\right. \ ,
\end{eqnarray}
where $\Delta_y$ and $S_y$ were defined in (\ref{dfDltaS}) and, in general, $\ell = \ell (\alpha)$, such that $\ell (0) = \ell_0$.

Here we want to ask whether or not  these two sets of vectors can be mapped to each other. More specifically, we ask if there exists a non-degenerate map $U$ such that
\begin{eqnarray}\label{U1}
\left\{
\begin{array}{l}
  U^{-1}{K}^{\mu}U = {K}_\star^{\mu} \\
  U|_{\alpha = 0} = \mathds{1} \\
\end{array}
\right.\ .
\end{eqnarray}
If this were to be the case, then the two theories, commutative and noncommutative, would essentially be equivalent, as one would be able to map all the solutions of one theory to the ones of the other, and it would be easy to see that this map would preserve the conformal structure of the $n$-point functions. The absence of such a map would tell us that the conformal behaviour of the noncommutative theory is really a non-trivial result. Below we will see that the latter is true.

Because the addition to $K^+$ of a term proportional to $\ell_0 e^{-x}\,\partial_y \equiv - e^{-2x} K^-$ does not spoil the $so(2,1)$ algebra,\footnote{This term appears naturally by defining the Killing vectors as Poisson vector fields (or as the adjoint action in the non-commutative case), leading to the correct equation of motion as the kernel of the Casimir operator. But it is \textit{not} needed to close the algebra, so it is more of a physical origin.} it makes sense to study the effect of this term separately. Towards this end introduce the ``shifted'' Killing vectors:
\begin{eqnarray}\label{shiftedKillings}
\left\{
\begin{array}{l}
  \tilde{K}^0=K^0=\partial_x\\
  \tilde{K}^- = K^-=  -\ell_0 e^{x}\,\partial_y \\
  \tilde{K}^+ =K^+ - \ell_0 e^{-x}\,\partial_y \equiv  \frac 1{\ell_0} e^{-x}\left(2 y\,\partial_x\,+y^2\partial_y \right)
\end{array}
\right. \ .
\end{eqnarray}

We will look for a similarity transformation that takes the ``shifted'' commutative generators (\ref{shiftedKillings}) to ``shifted'' non-commutative ones. Here by ``shifted'' non-commutative generators we mean the following
\begin{eqnarray}\label{NCshiftedKillings}
\left\{
\begin{array}{l}
  \tilde{K}_\star^0 = K_{\star}^0 \\
  \tilde{K}_\star^- = K_{\star}^- \\
  \tilde{K}_\star^+ = \frac 1{\ell} e^{-x}\left(2 y\,\partial_x S_y\,+(y^2 - \frac{\alpha^2}{4}\partial_x^2)\Delta_y \right) + const\times e^{-x} \Delta_y
\end{array}
\right. \ .
\end{eqnarray}
Note that adding $const\times e^{-x} \Delta_y $ to $K_\star^+$ also does not effect the algebra (so, (\ref{ncKillings}) corresponds to $const = \ell +\frac{\alpha^2}{4\ell}$). Of course, later we also will be interested in the effect of such a transformation (if exists) on the shift term.

To deal with the difference between $\ell_0$ in (\ref{shiftedKillings}) and $\ell$ in (\ref{NCshiftedKillings}), one can perform the trivial similarity transformation, $U_0 = \exp\left(-\ln(\frac{\ell}{\ell_0})\partial_x\right)$, which changes $\ell_0$ in (\ref{shiftedKillings}) to $\ell$ (It is important to note that this is not the case for the original, non-shifted, generators (\ref{Killings}) because $\ell_0$ enters $K^+$ not only via a common factor.) We will assume that this trivial similarity transformation has been done and will keep on using the same notation for (\ref{shiftedKillings}) but now with $\ell_0 \rightarrow \ell$.

Because in this paper we are interested in leading order perturbations in the noncommutativity parameter, we will look for the similarity transformation to the $\alpha^2$-order. To this order we have
\begin{eqnarray}\label{perturbNCshiftedKillings}
\left\{
\begin{array}{l}
  \tilde{K}_{\star}^0=\partial_x\\
  \tilde{K}_{\star}^- = \tilde{K}^- + \alpha^2\frac{\ell}{24}e^{x}\partial_x^3 + \mathcal{O}(\alpha^4) \\
  \tilde{K}_{\star}^+ = \tilde{K}^+ - \alpha^2  \frac 1{4\ell} e^{-x}\left( \partial_y \partial_x^2 + y\partial_y^2 \partial_x + \frac{1}{6}y^2 \partial_y^3 +\kappa\partial_y \right) + \mathcal{O}(\alpha^4)
\end{array}
\right. \ .
\end{eqnarray}
Here the term proportional to the \textit{unknown} constant $\kappa$ is exactly the possible shift term. More specifically, the constant in (\ref{NCshiftedKillings}) is equal to $-\frac{\alpha^2}{4\ell}\kappa$. We see that there is no $\alpha$-linear term, so it is natural to suggest the following expansion for the map $\tilde{U}$ from $\tilde{K}^\mu$ to $\tilde{K}^\mu_\star = \tilde{U}^{-1}\tilde{K}^\mu\tilde{U}$
\begin{eqnarray}\label{Uperturb}
\tilde{U} = \mathds{1} + \alpha^2 \mathcal{G}(x,\partial_x , y, \partial_y) + \mathcal{O}(\alpha^4) \ .
\end{eqnarray}
Then we have the following conditions on $\mathcal{G}$: $\alpha^2 [\tilde{K}^\mu , \mathcal{G}] = \tilde{K}_{\star}^\mu - \tilde{K}^\mu $ or in the components
\begin{eqnarray}
\left[\tilde{K}^0 , \mathcal{G}\right] &=& 0\ ,\label{cond1}\\
\left[\tilde{K}^- , \mathcal{G}\right] &=& \frac{\ell}{24}e^{x}\partial_x^3 \ ,\label{cond2}\\
\left[\tilde{K}^+ , \mathcal{G}\right] &=& - \frac{1}{4\ell} e^{-x}\left( \partial_y \partial_x^2 + y\partial_y^2 \partial_x + \frac{1}{6}y^2 \partial_y^3 +\kappa\partial_y \right) \ .\label{cond3}
\end{eqnarray}
Let us analyze these conditions one by one.

1) The condition (\ref{cond1}) requires that $\mathcal{G}$ does not depend on $x$.

2) The condition (\ref{cond2}) is $\left[e^x \partial_y , \mathcal{G}\right] = - \frac{1}{24}e^{x}\partial_x^3$. We note that there is a trivial solution to it: $\mathcal{G}_0 = -\frac{1}{24}y\partial_y^3$, so, writing $\mathcal{G}=\mathcal{G}_0 + \tilde{\mathcal{G}}$, this is equivalent to $\left[e^x \partial_y , \tilde{\mathcal{G}}\right] = 0$. Because the full form of the non-commutative Killing vectors depends on derivatives with respect to $x$ only up to $\partial_x^2$, it is possible to argue that $\tilde{\mathcal{G}}$ also does not involve terms with $\partial_x^{(k)}$ with $k\ge 3$, i.e. $\tilde{\mathcal{G}}$ takes the following form
\begin{eqnarray}\label{tilG}
\tilde{\mathcal{G}} = g_2 (y,\partial_y)\partial_x^2 + g_1 (y,\partial_y)\partial_x + g_0 (y,\partial_y)\ .
\end{eqnarray}
Taking into account the independence of $\partial_x^{(k)}$ for different $k$, after some trivial calculation, we arrive at the following most general form for the candidate for the infinitesimal similarity transformation:
\begin{eqnarray}\label{G1}
\mathcal{G} = -\frac{1}{24}y\partial_y^3 &+& g_2(\partial_y)\partial_x^2 + \Bigl(2y g_2(\partial_y)\partial_y + \tilde{g}_1(\partial_y)\Bigr)\partial_x + \nonumber \\
&+& 2 y^2 g_2(\partial_y)\partial_y^2 +y \Bigl( g_2(\partial_y) + \tilde{g}_1 (\partial_y)\Bigr)\partial_y + \tilde{g}_0(\partial_y) \ ,
\end{eqnarray}
where $g_2$ and $\tilde{g}_i$ are some arbitrary functions of the argument $\partial_y$.

3) Using the result of the previous step (\ref{G1}) in (\ref{cond3}) and requiring that the term proportional to $\partial_x^3$ is absent on the RHS of (\ref{cond3}) we immediately conclude that $g_2 (\partial_y)$ is actually a constant, i.e. the whole dependence on it drops out. Continuing to compare the coefficients of $\partial_x^{(k)}$ for different $k=0,1,2$, we arrive at the following result
\begin{eqnarray}\label{g10}
g_2=a\ ,\ \tilde{g}_1 = \frac{1}{16}\partial_y^2 + b\ ,\ \tilde{g}_0 = \frac{1}{2}\tilde{g}_1\ ,\ \kappa = -\frac{1}{4}\ ,
\end{eqnarray}
where $a$ and $b$ are some arbitrary constants, which do not contribute at this level. We still keep the dependence on $a$ and $b$ explicit to study the transformation of the shift term (see below).

This completes the prove of the perturbative (up to $\alpha^2$-terms) equivalence of $\tilde{K}^\mu$ and $\tilde{K}_{\star}^\mu$ (with the very precise form of the generated shift term)
\begin{align}\label{equivalence}
 & \tilde{U}^{-1}\tilde{K}^0 \tilde{U} = K_{\star}^0 = \partial_x \ ,\nonumber\\
 & \tilde{U}^{-1}\tilde{K}^- \tilde{U} = \tilde{K}_{\star}^- + \mathcal{O}(\alpha^4)\ ,\nonumber\\
 & \tilde{U}^{-1}\tilde{K}^+ \tilde{U} = \tilde{K}_{\star}^+ + \mathcal{O}(\alpha^4)= \frac 1{\ell} e^{-x}\left(2 y\,\partial_x S_x\,+(y^2 - \frac{\alpha^2}{4}\partial_x^2)\Delta_y \right) + \frac{\alpha^2}{16\ell} e^{-x}\Delta_y + \mathcal{O}(\alpha^4) \ ,
\end{align}
where
\begin{eqnarray}\label{Uperturb1}
\tilde{U} &=& \mathds{1} + \alpha^2 \mathcal{G}(\partial_x , y, \partial_y) + \mathcal{O}(\alpha^4) = \mathds{1} + \alpha^2\left( -\frac{1}{24}y\partial_y^3 + a(\partial_x^2 +2y^2 \partial_y^2 +2y\partial_y \partial_x + y \partial_y) +\right.\nonumber \\
&+&\left. \frac{1}{32}(2y\partial_y + 2\partial_x +1)\partial_y^2 + b(y\partial_y + \partial_x ) \right)+ \mathcal{O}(\alpha^4)\ .
\end{eqnarray}
Note, that as it was stressed above, (\ref{equivalence}) does not depend on the arbitrary $a$ and $b$.

For the future use, we consider a more general choice for $\tilde{g}_0$: $\tilde{g}_0 =\frac{1}{2}\left(1+\lambda\right)\tilde{g}_1$. Of course this will produce some extra terms on the right hand side, but we will see how they are cancelled by the shift term.\footnote{Changing $g_2$ or $\tilde{g}_1$ immediately will produce higher $x$-derivatives that will not be possible to compensate.} So, we have (including the contribution from $U_0$)
\begin{align}\label{Transfextra}
&\tilde{U} = \mathds{1} + \alpha^2 \mathcal{G}(\partial_x , y, \partial_y) + \mathcal{O}(\alpha^4) = \mathds{1} + \alpha^2\left( -\frac{1}{24}y\partial_y^3 + a(\partial_x^2 +2y^2 \partial_y^2 +2y\partial_y \partial_x + y \partial_y) + \right.\nonumber \\
\ \ \ &\qquad\qquad\qquad\qquad\qquad+ \left.\frac{1}{32}(2y\partial_y + 2\partial_x +1+\lambda)\partial_y^2 + b(y\partial_y + \partial_x ) -\frac{1}{\ell_0^2}\frac{\ell_1}{\ell_0}\partial_x\right)+ \mathcal{O}(\alpha^4)\ ,\nonumber\\
&\tilde{U}^{-1}\tilde{K}^0 \tilde{U} = K_{\star}^0 = \partial_x \ ,\nonumber\\
&\tilde{U}^{-1}\tilde{K}^- \tilde{U} = \tilde{K}_{\star}^- + \mathcal{O}(\alpha^4)\ ,\nonumber\\
&\tilde{U}^{-1}\tilde{K}^+ \tilde{U} = \tilde{K}_{\star}^+ + \mathcal{O}(\alpha^4)= \frac 1{\ell} e^{-x}\left(2 y\,\partial_x S_x\,+\Bigl(y^2 - \frac{\alpha^2}{4}\partial_x^2\Bigr)\Delta_y \right) + \nonumber \\
\ \ \ &\qquad\qquad\qquad\qquad\qquad+\frac{\alpha^2}{16\ell_0}(1-\lambda) e^{-x}\Delta_y -\frac{\alpha^2 \lambda}{8\ell_0}(\partial_y \partial_x +y\partial_y^2)+ \mathcal{O}(\alpha^4) \ .
\end{align}

The problem with the shift term, $const\times e^{-x}\partial_y$, is immediately clear from the fact that neither expression in front of the constants $a$ and $b$ in (\ref{Transfextra}) commutes with this term. So, as the consequence, we will produce terms explicitly depending on these constants. It is easy to obtain the perturbative form of the transformation of the shift term (we expand $\ell = \ell_0 + \frac{\alpha^2 }{\ell_0^2} \ell_1 + \mathcal{O}(\frac{\alpha^4 }{\ell_0^4})$)
\begin{align}\label{similarityshift}
U^{-1}U_0^{-1} (\ell_0 e^{-x}\,\partial_y ) U_0 U = \ell e^{-x}\Delta_y + \alpha^2 \ell_0 e^{-x}\left(4a(\partial_x + y\partial_y ) +2b - \frac{2\ell_1}{\ell_0^3} +\frac{1}{8}\partial_y^2\right)\partial_y + \mathcal{O}(\alpha^4)\ .
\end{align}
While the first term has a correct form (which, of course, remains correct after the expansion in $\alpha$ is done), the rest presents a correction (the difference between $\ell$ and $\ell_0$ is of the next order in $\alpha$.)

Combining (\ref{Transfextra}) and (\ref{similarityshift}) one can easily see that the choice $\lambda=32a\ell_0^2$ and $b-a = \frac{1}{\ell_0^2}\left(\frac{\ell_1}{\ell_0}+\frac{3}{32}\right)$ (the separate values of $a$ and $b$ turn out to be irrelevant) almost does the mapping between the two sets of Killing vectors, (\ref{Killings}) and (\ref{ncKillings}) (we return to the ``untilded'' notation for $U$, because this is a map between $K^{\mu}$ and $K^{\mu}_\star$ as in (\ref{U1}))
\begin{align}\label{Transffinal}
&U = \mathds{1} + \alpha^2 G(\partial_x , y, \partial_y) + \mathcal{O}(\alpha^4) = \nonumber \\
\ \ \  &\;\;\;=\mathds{1} + \alpha^2\left( \frac{1}{96}(2y\partial_y + 6\partial_x +3)\partial_y^2 + \frac{1}{\ell_0^2}\Bigl(\frac{\ell_1}{\ell_0}+\frac{3}{32}\Bigr) (y\partial_y + \partial_x)-\frac{1}{\ell_0^2}\frac{\ell_1}{\ell_0}\partial_x \right)+ \mathcal{O}(\alpha^4)\ ,\nonumber\\
 & U^{-1}{K}^0 U = K_{\star}^0 \ ,\nonumber\\
 & U^{-1}{K}^- U = {K}_{\star}^- + \mathcal{O}(\alpha^4) \ ,\nonumber\\
 & U^{-1}{K}^+ U = {K}_{\star}^+ + \alpha^2 \frac{\ell_0}{8}e^{-x}\partial_y^3 + \mathcal{O}(\alpha^4) \ .
\end{align}

To conclude, we can \textit{almost} map the commutative Killing vectors to the noncommutative ones. The obstruction is the shift term, the presence of which leads to the appearance of the extra $\partial_y^3$-term. This is not only the proof of the ``non-triviality'' of $ncAdS_2$ but also serves as very convenient technical tool to simplify the perturbative analysis of the sections \ref{SectionNCfree} and \ref{SectionNCint}.

\section{Noncommutative boundary term}\label{AppendixBoundary}
\setcounter{equation}{0}

Here we want to show that the commutative expression for the on-shell action (\ref{fortee7}) is valid in the non-commutative case to \textit{all orders} in $\alpha$. Towards this end, let us re-write (\ref{ncosbndryactn}) in terms of the non-commutative Killing vectors (\ref{ncKillings})
\begin{align}\label{boundary}
S_{nc}^{bdy}[\Phi]= -\frac 1{2\ell\alpha^2} \int dxdy\,[{\cal X}^\mu,\Phi\star [{\cal X}_\mu,\Phi]_\star]_\star = \frac{1}{2\ell}\iint\limits_{\mathbb{R}^2} dx dy K^\mu_{\star} \left(\Phi\star K_{*\mu}  \Phi \right) \ ,
\end{align}
where $K^\mu_{\star}$ are defined as $\alpha K^\mu_{\star} \Phi := i[\mathcal{X}^\mu , \Phi]_\star$.

In two dimensions, the Stokes theorem takes the form ($\omega = \omega_\mu dx^{\mu}$ is an arbitrary 1-form)
\begin{align}
\iint\limits_{V} dx dy (\partial_x \omega_y - \partial_y \omega_x) = \int\limits_{\partial V} \omega_x dx + \omega_y dy \ .
\end{align}
Because the boundary of our space is located at $z=0$ it is convenient to pass to Fefferman-Graham coordinates (\ref{FGsymbol}). Then $z=const$ corresponds to $x=const$ with $dy = \frac{\ell}{z}dt$ and for our case the Stokes formula takes the form
\begin{align}\label{Stokes}
\iint\limits_{\mathbb{R}^2} dx dy (\partial_x \omega_y - \partial_y \omega_x) = \int\limits_{-\infty}^\infty \left[\frac{\ell}{z}\omega_y\right]_{z=0} dt \ .
\end{align}
This means that while studying the integrand of (\ref{boundary}) we need to keep track only of the term of the form $\partial_x (\cdots)$. Also, because $\frac{\ell}{z}\omega_y$ is evaluated at $z=0$, we only need to keep terms in $\omega_y$ up to $\mathcal{O}(z^2)$. This will allow us to arrive at the exact result. Using
\begin{align}
\left\{\begin{array}{l}
  \left[\Delta_y , y\right] = S_y  \\
  \left[S_y , y\right] = -\frac{\alpha^2}{4}\Delta_y
\end{array}\right.
\ \Rightarrow\ [\Delta_y , y^2] = 2y S_y -\frac{\alpha^2}{4}\Delta_y \ ,
\end{align}
where $\Delta_y$ and $S_y$ are defined in (\ref{dfDltaS}), the Killings (\ref{ncKillings}) take the form
\begin{eqnarray}\label{ncKillings1}
\left\{
\begin{array}{l}
  K_{\star}^0=\partial_x\equiv K^0 \\
  K_{\star}^-= -\Delta_y\ell\,e^{x} \\
  K_{\star}^+= \partial_x\frac 2{\ell} e^{-x} y\, S_y+\Delta_y \frac 1{\ell} e^{-x}\left(y^2 + \ell^2 +\frac{\alpha^2}{2} - \frac{\alpha^2}{4}\partial_x^2 \right)
\end{array}
\right. \ ,
\end{eqnarray}
where we moved all the relevant derivatives to the left (note that $\Delta_y$ has the form $\partial_y (\cdots)$). Then we have
\begin{align}
K^\mu_{\star} \left(\Phi\star K_{*\mu}  \Phi \right) &= K^0_{\star}\left(\Phi\star K_{\star}^0  \Phi \right) - \frac{1}{2}K^+_{\star}\left(\Phi\star K_{\star}^-  \Phi \right) - \frac{1}{2}K^-_{\star}\left(\Phi\star K_{\star}^+  \Phi \right) = \nonumber\\
&=\partial_x \left(\Phi\star K_{\star}^0  \Phi   - \frac{1}{\ell} e^{-x} y\, S_y \left(\Phi\star K_{\star}^-  \Phi \right) \right) - \partial_y (\cdots) \ .
\end{align}
So we need to find the form, up to $\mathcal{O}(z^2)$, of the following expression
\begin{align}
\Phi\star K_{\star}^0  \Phi   - \frac{1}{\ell} e^{-x} y\, S_y \left(\Phi\star K_{\star}^-  \Phi \right) \equiv -\Phi\star (z\partial_z +t\partial_t) \Phi   + t\, S_t \left(\Phi\star \Bigl(\frac{\ell}{z}\Delta_t  \Phi \Bigr) \right) \ ,
\end{align}
where we passed to FG coordinates and $S_t = \cos\left( \frac{\alpha}{2\ell}z\partial_t \right)$ and $\Delta_t = \sin\frac{2}{\alpha}\left( \frac{\alpha}{2\ell}z\partial_t \right)$. Using these coordinates, the derivatives are given by
\begin{align}
\left\{\begin{array}{l}
  \partial_x = -z\partial_z - t\partial_t  \\
  \partial_y = \frac{z}{\ell}\partial_t
\end{array}\right.
\end{align}
it is obvious that the star-product (\ref{dffstrprd}),
\begin{align}\label{starprod}
\star = \sum\limits_{k=0}^\infty \frac{1}{k!}\left(\frac{i\alpha}{2}\right)^k \epsilon^{i_1 j_1}\cdots \epsilon^{i_k j_k}\overleftarrow{\partial_{i_1}\cdots\partial_{i_k}}\overrightarrow{\partial_{j_1}\cdots\partial_{j_k}}\ ,\ (x^1,x^2):=(x,y)
\end{align}
cannot lower the degree of $z$. Moreover, every time we apply the derivative $\partial_y$, we raise the degree of $z$ by 1. This, combined with the fact that
$$
\frac{\ell}{z}\Delta_t = \partial_t + \mathcal{O}(z^2)\ ,\ \ S_t = 1 + \mathcal{O}(z^2)
$$
allows us to write
\begin{align}
-\Phi\star (z\partial_z +t\partial_t) \Phi   + t\, S_t \left(\Phi\star \left(\frac{\ell}{z}\Delta_t  \Phi \right) \right) = -\Phi\star (z\partial_z +t\partial_t) \Phi   + t\, \left(\Phi\star \partial_t  \Phi \right) + \mathcal{O}(z^2) \ .
\end{align}
Using the explicit expression for the star-product (\ref{starprod}), we see that it actually starts with the terms of the order of $z^2$ (also, see the discussion in Appendix \ref{SectionxyFG})
\begin{align}\label{starprod1}
 \star = 1 + \frac{i\alpha}{2\ell} \left( \overleftarrow{\partial_t} z^2 \overrightarrow{\partial_z} - \overleftarrow{\partial_z} z^2 \overrightarrow{\partial_t} \right) + \sum\limits_{k=2}^\infty \frac{1}{k!}\left(\frac{i\alpha}{2}\right)^k \epsilon^{i_1 j_1}\cdots \epsilon^{i_k j_k}\overleftarrow{\partial_{i_1}\cdots\partial_{i_k}}\overrightarrow{\partial_{j_1}\cdots\partial_{j_k}}\ ,
\end{align}
where the remaining sum is at least of the order of $\mathcal{O}(z^2)$. This finally allows us to write
\begin{align}
-\Phi\star (z\partial_z +t\partial_t) \Phi   + t\, S_t \left(\Phi\star \left(\frac{\ell}{z}\Delta_t  \Phi \right) \right) = -z \Phi \partial_z \Phi + \mathcal{O}(z^2) \ .
\end{align}
Multiplying this by $\frac{\ell}{z}$, evaluating at $z=0$, plugging into the boundary part of the action (\ref{boundary}) and taking into account (\ref{Stokes}), we get (\ref{fortee7}) as an exact result.

\section*{Acknowledgements}

We are very grateful to M. Kaminski, C. Uhlemann and J. Wu for valuable discussions.

\bibliography{actiondiracbib}

 \bibliographystyle{utphys}
\end{document}